\documentclass[a4paper]{article}
\usepackage[a4paper,margin=2.5cm,top=2.7cm]{geometry}

\usepackage{amsmath}
\usepackage{amssymb}
\usepackage{bm}
\usepackage{graphicx}
\usepackage[hidelinks]{hyperref}
\usepackage{cleveref}
\crefname{equation}{Equation}{Equations}
\usepackage{subcaption}
\usepackage{tikz}
\usetikzlibrary{positioning}
\usetikzlibrary{calc}
\usepackage{arydshln}
\usetikzlibrary{matrix, positioning, calc}
\usepackage{comment}
\usepackage{float}
\usepackage{placeins}

\usepackage[compress]{natbib}
\usepackage{doi}

\usepackage{authblk}
\title{Investigating the mechanism by which finite-size heavy particles are entrained in turbulent open channel flow over a smooth surface}
\author{Tatia Bzikadze$^{1,*}$, Markus Weyrauch$^{1,\dagger}$ and Markus Uhlmann$^{1,\ddagger}$}
\affil{$^{1}$Institute for Water and Environment - Numerical Fluid Mechanics Group, Karlsruhe Institute of Technology, 76131 Karlsruhe, Germany}
\affil{Email: $^{*}$tatia.bzikadze@kit.edu, $^\dagger$markus.scherer@kit.edu, $^\ddagger$markus.uhlmann@kit.edu}
\date{}

\begin{document}

\setlength{\affilsep}{1em}        
\renewcommand{\Affilfont}{\small} 

\maketitle





\begin{abstract}
The dynamics of entrainment of finite-size heavy particles in a turbulent open channel flow over a smooth surface are analyzed. Three types of simulations, namely with freely moving, rotation-constrained, and spanwise-motion-constrained particles, were conducted using particle-resolved direct numerical simulations. With the aid of a relative velocity suitably defined in the vicinity of the finite-size particle, we decompose the hydrodynamic force into drag and lift contributions and evaluate the local wall-normal shear rate around the particles. By means of coherent structure eduction techniques, we investigate flow structures before and during lift-off events. Rotation-constrained simulations revealed the insignificance of particle rotation in the entrainment mechanism. Spanwise-motion-constrained simulations revealed the importance of particle location with respect to flow structures with apparent changes in entrainment frequency, duration of the entrainment process, wall-normal shear around the particles, and distance to the nearest vortical structures during lift-off. The contribution of lift to the wall-normal force is found to be responsible for the initiation of particle entrainment, which is induced by a high-shear event associated with fast-moving fluid. The presence of quasi-streamwise vortices is shown to be an important ingredient for the entrainment of particles into the bulk flow. The results show that, at marginal Shields number values, a high wall-normal shear rate and the proximity of an intense quasi-streamwise vortex are essential elements of the entrainment mechanism.

\end{abstract}

\section{Introduction}
\label{sec:introduction}

Within climate and hydrological systems, particle entrainment is an important component in the transport and accumulation of particles in water. The suspension of solid particles is the major factor affecting water quality \citep{BILOTTA20082849}. Microplastics resuspended and transported by rivers into the oceans are a major contributor to pollution \citep{rochmanpollution}. Therefore, understanding what induces particle vertical and horizontal motion is relevant in many environmental processes. Research has consistently shown that particle motion varies under different flow conditions and particle parameters. Consequently, the mechanisms underlying particle entrainment remain only partially understood. 

Heavy particles, which are introduced into a horizontal open channel flow, tend to settle to the bottom wall and can accumulate in low-speed streaks. This phenomenon is now well established based on a variety of experimental and numerical studies \citep{ninto1996experiments,suzuki2000simultaneous,kidanemariam2013direct,pestana2020can,peng2024preferential}. Deposited particles can detach from the bottom wall and resuspend through the action of fluid flow. 
The study of a threshold for incipient motion, the parameter point at which the particle starts to move, can be traced back to the work of \citet{shields:36}. He evaluated the forces acting on individual grains, postulating a criterion relating the wall shear stress and the particles submerged weight.

The hydrodynamic forces can be evaluated in numerical simulations when the instantaneous flow field around the finite-size particle is fully resolved. On the contrary, for point-particle simulations, when the particle size is smaller than the grid size, the hydrodynamic forces are estimated by approximate formulas. The Maxey–Riley–Gatignol equation \citep{maxey1983equation,gatignol1983faxen} describes the hydrodynamic forces acting on a single isolated particle and includes buoyancy, added mass, quasi-steady drag, Basset history forces, and the force by the undisturbed velocity field. In many cases, lift force is also included in the equation \citep{m2016point}. In the following discussion, we are going to focus on the quasi-steady drag and lift forces. \citet{loth2008lift} conducted a comprehensive review of different mechanisms leading to the generation of lift force. \citet{saffman1965lift} introduced the Saffman lift force in a uniform linear shear flow for a non-rotating particle and small particle Reynolds numbers. \citet{mclaughlin1991inertial} and \citet{mei1994note} extended shear-induced lift analysis for higher particle Reynolds numbers. Moreover, rotation of the particle is found to induce an upward lift force. \citet{rubinow1961transverse} proposed the Robins-Magnus lift force expression for spin-induced lift for a rotating particle in a uniform flow and small particle Reynolds numbers. \citet{bagchi2002effect} suggested a linear combination of these forces to study particle lift with both spin and linear shear. \citet{shi2019lift} provided a review of existing results from analytical, numerical, and experimental studies. However, these studies focus only on particles in laminar, unbounded flows. \citet{zeng2005wall,zeng2009forces}, \citet{lee2010drag}, and \citet{shi2021drag} 
conducted fully resolved simulations of the flow around a rigid sphere in a wall-bounded fluid at rest and in a wall-bounded linear shear flow and derived empirical expressions for lift and drag forces based on the results for non-rotating and rotating spheres. Improving the empirical expressions for quasi-steady drag and lift remains an active area of research.

In particle-laden turbulent flows, the hydrodynamic forces exhibit fluctuations due to the unsteady, chaotic nature of the flow. It has previously been observed that near-bed turbulent structures that interact with particles can cause particle entrainment. \citet{soldati2009physics} investigated mechanisms responsible for the deposition and entrainment of point-particles in turbulent dispersed flow using direct numerical simulations. They showed that sweeps tend to be correlated with particles moving toward the wall, while ejections entrain particles toward the outer flow. The experimental study of \citet{ninto1996experiments} shows that extreme flow ejection events are related to particle entrainment. 

\citet{kidanemariam2013direct} conducted particle-resolved direct numerical simulations (PR-DNS) of horizontal turbulent open channel flow. The particle diameter and the Shields number were chosen such that, after settling, the particles essentially remain in contact with the wall during the entire simulation. Through conditional averaging, they showed that spanwise particle motion is caused by nearby quasi-streamwise vortices. This explains the migration of the particles into the low-speed streaks. \citet{dwivedipressure,dwivedi2011flow} investigated the role and effect of coherent structures on the hydrodynamic forces responsible for the inception of particle motion by an experimental study on a rough bed. Sweep events were found to lift a completely shielded particle from its cavity, with lift being the dominant mechanism, while drag, on the contrary, was found to be the dominant mechanism for exposed particles, causing them to roll over the neighboring particle. Pressure gradients coincided with the occurrence of a sweep event, indicating the importance of pressure gradients during the inception of particle motion. \citet{vowinckel2016entrainment} conducted PR-DNS of a turbulent open channel flow. The entrainment of an individual particle out of its pocket was studied after mobile particles were released over a sediment bed that is composed of fixed particles at the wall. These simulations revealed that for the resting particle to be entrained, the collision with an already entrained particle is important. They showed that, after such a slight initial displacement of the particle from its pocket, a strong sweep event is present during the entrainment to completely dislodge the particle from its pocket. Furthermore, they reported that, near the entrained particle, two regions of counter-rotating fluid can be identified during the entrainment event. Further experimental studies of sediment entrainment from a bed of spherical particles were conducted by \citet{cameron2020entrainment} at low and intermediate particle protrusions. They stated that particle entrainment is related to very large-scale motion and pressure fluctuations, and emphasized the importance of the duration of force fluctuations to completely entrain the particle from its pocket. However, the role of lift forces has not been investigated by those authors. \citet{yousefi2020single} studied the dynamics of a single particle in a turbulent open channel flow over a fixed porous bed for different Shields numbers. They showed that particle motion coincides with the intense streamwise velocity fluctuations associated with the sweep events. Additionally, sweeps present at the moment of particle motion are related to the tail of a nearby hairpin-like vortex. The presence of the passage of the hairpin vortex was consistent with the investigated local pressure distribution at the surface of the particle. 

As summarized above, most previous studies have examined the dynamics of particle entrainment over rough surfaces, where the question addressed is how the particle, which is initially geometrically constrained by the shape of the bed, lifts off from its pocket. In this study, we investigate the dynamics of individual finite-size heavy particles entrained in a pressure-driven, turbulent open channel flow over a smooth wall by performing PR-DNS. The physical parameters are chosen to match those of \citet{kidanemariam2013direct}, however, at a different Shields number. While the particles in their study remained near the wall, here the Shields number is selected slightly above the critical threshold to allow resuspension into the flow, but small enough that only intense coherent structures can induce particle lift-off. To further examine the influence of particle motion, constrained simulations are performed where rotation and spanwise motion are suppressed. The aim is to identify individual lift-off events, analyze the hydrodynamic forces acting on the particle during its entrainment, and determine the flow structures responsible for particle lift-off.  

The manuscript is organized as follows. Section~\ref{sec:flow_setup} describes the governing equations and numerical methods, as well as the flow and particle parameters. In \autoref{sec:analysis}, we define the quantities and describe the methodologies used to analyze the dynamics of entrainment events. Section~\ref{sec:result} presents the results obtained using these methods. A summary of the results and concluding remarks is given in \autoref{sec:conclusion}.

\section{Flow setup and methods}
\label{sec:flow_setup}
\subsection{Equations of motion}
\label{sec:motion_equations}

\begin{figure}
    \centering
    \includegraphics[width=0.8\linewidth]{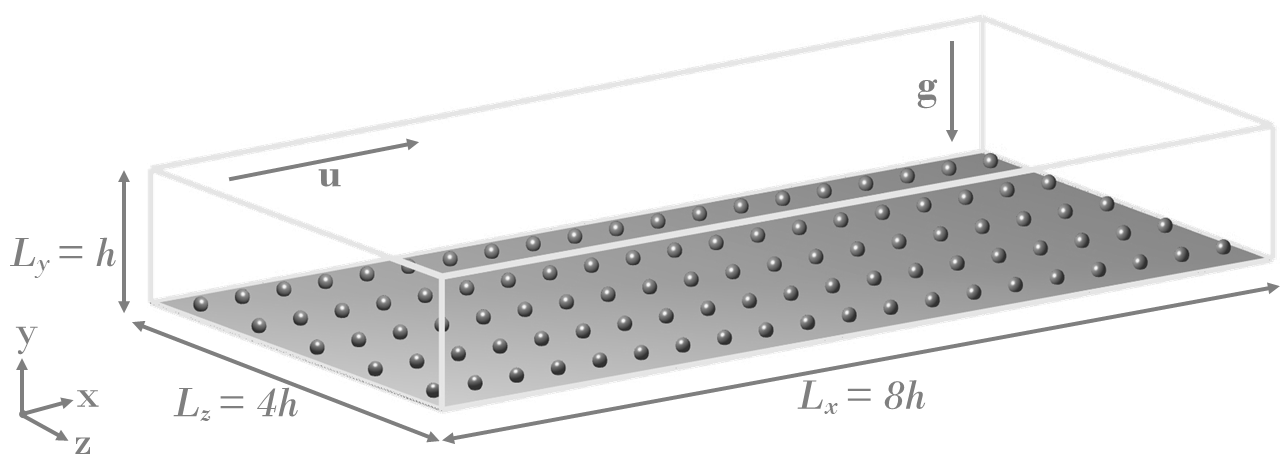}
    \caption{Flow configuration. The streamwise and spanwise directions are periodic. Gravity is directed in the negative wall-normal direction. Particles are shown at their initial locations.}
    \label{fig:configuration}
\end{figure}

The motion of the carrier fluid is governed by the incompressible Navier-Stokes equations,

\begin{align}
    \nabla \cdot \textbf{u}_\mathrm{f} &= 0, 
    \label{eq:NS1}\\ 
    \frac{\partial \textbf{u}_\mathrm{
f}}{\partial t} + (\textbf{u}_\mathrm{f} \cdot \nabla)\textbf{u}_\mathrm{f} &= - \frac{1}{\rho_\mathrm{f}} \nabla p + \nu_\mathrm{f} \nabla ^2 \textbf{u}_\mathrm{f}, 
    \label{eq:NS2} 
\end{align}
where $\textbf{u}_\mathrm{f} = (u_\mathrm{f},v_\mathrm{f},w_\mathrm{f})$ denotes the velocity of the fluid, $p$ the pressure, $\rho_\mathrm{f}$ the density of the fluid and $\nu_\mathrm{f}$ the viscosity of the fluid. $t$ is the time, and as shown in \autoref{fig:configuration}, the streamwise, wall-normal, and spanwise directions are $x$, $y$, and $z$, respectively. The domain is periodic in the streamwise and spanwise directions, while in the wall-normal direction, a no-slip boundary condition is applied at the bottom wall and a free-slip boundary condition at the top. 

Spherical particles are studied under the action of gravity $\mathbf{g}$, which is directed in the negative wall-normal direction. The dynamics of particles are governed by the Newton-Euler equations for the linear and angular momentum of a rigid body, and the equations for the $i$th particle are given by

\begin{align}
    \rho_\mathrm{p} V_\mathrm{p} \frac{d \mathbf{u}_\mathrm{p}^{(i)}}{dt} &= \rho_\mathrm{f} \oint_{\partial I^{(i)}} \boldsymbol{\tau} \cdot \mathbf{n}_{I^{(i)}} \, d\sigma + (\rho_\mathrm{p} - \rho_\mathrm{f}) V_\mathrm{p} \mathbf{g} + \mathbf{F}_{\mathrm{c}}^{(i)}, \label{eq:particle-momentum} 
    \quad \forall i = 1, \dots, N_\mathrm{p}, \\
    I_\mathrm{p} \frac{d \boldsymbol{\omega}_\mathrm{p}^{(i)}}{dt} &= \rho_\mathrm{f} \oint_{\partial I} \mathbf{r}^{(i)} \times (\boldsymbol{\tau} \cdot \mathbf{n}_{I^{(i)}}) \, d\sigma, 
    \quad \forall i = 1, \dots, N_\mathrm{p},
    \label{eq:particle-angular}
\end{align}
where $\textbf{u}_{\mathrm{p}} = (u_{\mathrm{p}},v_{\mathrm{p}},w_{\mathrm{p}})$ is the particle linear velocity, $\boldsymbol{\omega}_{\mathrm{p}} = (\omega_{{\mathrm{p}},x},\omega_{{\mathrm{p}},y},\omega_{{\mathrm{p}},z})$ the particle angular velocity, $\rho_{\mathrm{p}}$ and $V_{\mathrm{p}}$ the density and volume of the particle, respectively, and $I_{\mathrm{p}}$ the moment of inertia. $\partial I$ denotes the fluid-solid interface, while $\mathbf{n}_I$ is the outward-pointing normal vector on $\partial I$ and $\mathbf{r}$ the position vector with respect to the particle's center $\mathbf{r} = \mathbf{x} - \mathbf{x}_{\mathrm{p}}$, $\mathbf{x}_{\mathrm{p}} = (x_{\mathrm{p}},y_{\mathrm{p}},z_{\mathrm{p}})$ being the particle position vector. The hydrodynamic stress tensor is given by

\begin{align}
\boldsymbol{\tau} = - \frac{1}{\rho_{\mathrm{f}}} p \mathbf{I} + 2 \nu_{\mathrm{f}} \left( \nabla \mathbf{u}_{\mathrm{f}} + (\nabla \mathbf{u}_{\mathrm{f}})^\top \right),
\end{align}
with $\mathbf{I}$ representing the identity tensor. $N_{\mathrm{p}}$ is the total number of particles in a computational domain. $\mathbf{F}_{\mathrm{c}}$ refers to the solid contact force acting on the $i$th particle due to collision with either another particle or with the solid wall. Note that with the simple repulsion model used here (cf. \S~\ref{sec:numerical_method}), which does not involve tangential forces, no torque is generated at contact. 

\subsection{Numerical Method}
\label{sec:numerical_method}

The formulation of an immersed boundary method (IBM) developed by \citet{uhlmann2005immersed} is utilized to solve \eqref{eq:NS1} to \eqref{eq:particle-angular}, due to the presence of heavy, finite-size particles in the flow. In this approach, the direct forcing term is used to fulfill the no-slip boundary condition on the surface of the immersed body and is added on the right-hand side of \eqref{eq:NS2}. The computational grid is discretized using a non-body-conformal grid method, where Lagrangian points represent the surface of the immersed body and fixed Eulerian points represent the flow field. The fluid velocity is solved in the Eulerian frame and interpolated to the Lagrangian frame. The forcing term is then calculated in the Lagrangian frame to calculate the velocity that satisfies the no-slip condition at each Lagrangian point. Finally, the forcing term is spread onto the Eulerian grid, and the force is obtained. The regularized delta function introduced by \citet{peskin2002immersed} is used to perform the interpolation of the preliminary velocity and the force between the two frames. The Navier-Stokes equations are solved with a standard fractional step method, time integration is done with a three-step low-storage Runge-Kutta method for the nonlinear part, and the Crank-Nicolson scheme for the viscous term. For the space integration, a second-order, central finite-difference approach on a staggered grid is used.

Due to the action of gravitational acceleration and hydrodynamic forces, particles in the simulations can approach the confining boundary walls or other particles closely. To avoid overlapping of the regions occupied by the rigid bodies, a short-range repulsive force is added to the right-hand side of \eqref{eq:particle-momentum}, and it is activated when the distance between the particle and another body is smaller than a predefined force range. The IBM proposed by \citet{uhlmann2005immersed} uses the collision model introduced by \citet{glowinski1999distributed}, and the force range is set to two grid spacings as in \citet{kidanemariam2013direct}.

\subsection{Flow and particle parameters}
\label{sec:flow_particle_params}

The particle-to-fluid density ratio $\rho_\mathrm{p} / \rho_\mathrm{f}$ is set equal to 1.7, and the diameter of the particle $D_\mathrm{p}$ compared to the depth of the channel $h$ gives $D_\mathrm{p}/h = 10/256$. The particle diameter scaled in wall units reads $D^+_\mathrm{p} = D_\mathrm{p} / \delta_\nu = 7.21$, where $\delta_\nu$ is the viscous length scale defined as $\delta_\nu = \nu_\mathrm{f} / u_\tau$. The friction velocity $u_\tau$ is given by $u_\tau = \sqrt{\langle \tau_\mathrm{w} \rangle / \rho_\mathrm{f}}$, where $\langle \tau_\mathrm{w} \rangle$ is the wall shear stress averaged in time and over $x$ and $z$ at the bottom wall $y=0$. Therefore, the friction Reynolds number reads $Re_\tau = u_\tau h / \nu_\mathrm{f} = 184$. The bulk Reynolds number of the flow is $Re_\mathrm{b} = u_\mathrm{b} h / \nu_\mathrm{f} = 2870$ with $u_\mathrm{b}$ being the bulk velocity given by $u_\mathrm{b} = \frac{1}{h}\int_0^{h} \langle u_\mathrm{f} \rangle \, dy$, and the bulk fluid time scale is given by $t_\mathrm{b} = h / u_{\mathrm{b}}$. The resulting value of the Shields number, which is defined as 

\begin{align}
\theta = \frac{u_\tau^2}{\left( \frac{\rho_\mathrm{p}}{\rho_\mathrm{f}} - 1 \right) \mathbf{g} D_\mathrm{p}},
\end{align}
is then 0.22. In the study of \citet{kidanemariam2013direct}, a Shields number of 0.19 was found to be close to the critical value, as nearly all of the mobile particles remained concentrated near the wall. In order to generate a reasonable number of lift-off events during a limited temporal interval, we therefore selected a slightly higher Shields number in the present work. Note that the definition of a critical value is highly dependent on the way by which it is being determined (e.g. the precise definition of what lift-off means, the temporal horizon that has been examined, the size of the observed spatial volume, etc.). As an example, in the experimental study of \citet{https://doi.org/10.1046/j.1365-3091.2003.00551.x}, a critical Shields value of approximately 0.1 has been reported for the entrainment of particles from a smooth wall, at $D^+_\mathrm{p} \approx 4$, solid-to-fluid density ratio 2.65, and featuring a significantly larger particle concentration.

The size of the computational domain in all simulations is $[L_x,L_y,L_z] = [8h,1h,4h]$ and it is discretized with a uniform isotropic grid with widths $\Delta x = \Delta y = \Delta z = h/256$, that gives $D
_p / \Delta x = 10$ and $\Delta x^+ = \Delta y^+ = \Delta z^+ = 0.7$. $N_\mathrm{p} = 100$ particles are released in the domain to study the particle entrainment. The initial locations are chosen such that no direct interaction between particles occurs initially (cf. \autoref{fig:configuration}). The global solid volume fraction given by $\Phi_\mathrm{s} = N_\mathrm{p}V_\mathrm{p} / V_\mathrm{tot}$, where $V_\mathrm{tot}$ is the total volume enclosed by the computational domain, is equal to $10^{-4}$. Flow and particle parameters are summarized in \autoref{tab:quantities_table}.

\begin{table}[t]
\centering
\caption{Flow and particle parameters for the simulations. $Re_\mathrm{b}$ and $Re_\tau$ denote the bulk and friction Reynolds numbers, respectively, $\rho_\mathrm{p} / \rho_\mathrm{f}$ is the particle-to-fluid density ratio, $D_\mathrm{p}/h$ is the relative particle size, $D^+_\mathrm{p}$ is the particle diameter in wall units, $\theta$ is the Shields parameter, $\Phi_\mathrm{s}$ is the global solid volume fraction, and $N_\mathrm{p}$ is the number of particles in the domain.}
\label{tab:quantities_table}
\begin{tabular}[t]{cccccccc}
\hline
$Re_\mathrm{b}$&$Re_\tau$&$\rho_\mathrm{p} / \rho_\mathrm{f}$&$D_\mathrm{p}/h$&$D^+_\mathrm{p}$&$\theta$&$\Phi_\mathrm{s}$&$N_\mathrm{p}$\\
\hline
2870&184&1.7&10/256&7.21&0.22&$10^{-4}$&100\\
\hline
\end{tabular}
\end{table}

\subsection{Modified simulations}
\label{sec:modified_sims}

Particle linear and rotational motion \cref{eq:particle-angular,eq:particle-momentum} include the hydrodynamic force and torque acting on the particle. The effect of particle rotation on particle motion, and especially the entrainment process, has been the object of significant interest. It has been shown that particle rotation can induce lift \citep{rubinow1961transverse,loth2008lift}. In order to evaluate the effect of particle rotation on heavy, finite-size particle entrainment in wall-bounded turbulent flows, we conduct constrained simulations in addition to the simulations with freely moving particles, where the quantity of interest, in this case, angular velocity, is kept equal to zero. 

On the other hand, in wall-bounded turbulent flows, particles that are settled to the bottom wall tend to move in the spanwise direction into the low-speed streaks by the action of quasi-streamwise vortices, and particles tend to reside preferentially in the low-speed streaks for relatively long time intervals, thereby mostly sampling low-speed flow regions. Consequently, the third type of simulations we conducted consists of simulations in which particle spanwise motion is constrained, meaning that the initial particle spanwise location is fixed throughout the entire time interval of the simulation, while the particles remain free to rotate and translate in the $x$ and $y$ directions. Fixing the spanwise location will not allow particles to migrate into the low-speed streaks, which can increase the frequency of particles sampling other flow structures. The method of constrained simulations has previously been employed by \citet{patankar2001lift}, \citet{joseph2002slip}, and \citet{pestana2020can}.
Throughout the present text, the superscripts UN, RC, and ZC correspond to unconstrained, non-rotating, and spanwise-constrained cases, respectively.

\section{Analysis of entrainment events}
\label{sec:analysis}
\subsection{How to determine the velocity seen by a finite-size particle?}
\label{sec:how_to_relvel}

The relative velocity of the particle with respect to the surrounding fluid is an important quantity in particle-laden turbulent flows. It is the driver for the forces acting on the particle. The hydrodynamic forces acting on the particle depend on the local flow direction relative to the particle. Using the relative velocity provides a more precise direction of the local flow field and a consistent definition of drag and lift. The relative velocity is used to calculate the Reynolds number of the particles $Re_\mathrm{p} = D_\mathrm{p} |\mathbf{u}_{\text{rel}}^S| / \nu_\mathrm{f}$, where $\mathbf{u}_{\text{rel}}^S$ is defined below, and to calculate the drag and lift forces. However, what is taken as the surrounding fluid in the relative velocity calculation is not uniquely defined. One approach is to calculate the undisturbed fluid velocity at the particle location $\mathbf{x}_\mathrm{p}$. In point-particle models, this value is readily available when the simulations are one-way coupled, and the particle does not affect the flow. For finite-size particles, one approach is to define a concentric shell around the particle and take the average fluid velocity over the shell as the surrounding fluid velocity for the relative velocity calculation. This approach has been used by \citet{kidanemariam2013direct}, \citet{cisse2013slipping}, \citet{uhlmann2014sedimentation} and others. Furthermore, \citet{li2019fully} gave different definitions of relative velocity, including the undisturbed flow velocity at the particle center, the undisturbed flow velocity averaged over the particle surface, and the velocity taken at a point that lies one diameter upstream of the particle. However, to obtain an undisturbed flow at the particle center requires additional DNS simulations with the same flow configuration, but without the particle of interest. 

\begin{figure}
\centering
         \begin{minipage}{.46\linewidth} 
		       \raggedright{(\textit{a})}
           \includegraphics[width=\linewidth]
           {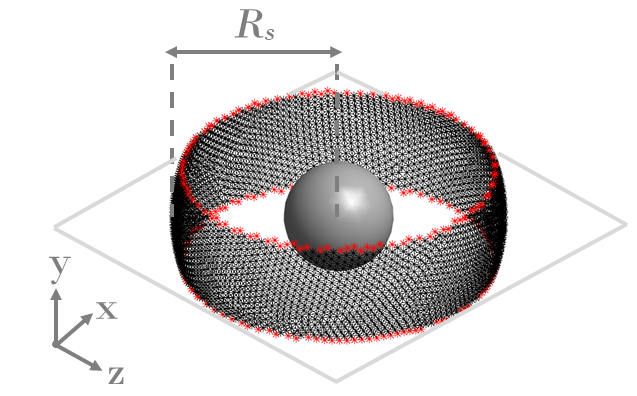}
         \end{minipage}
         %
         %
         \begin{minipage}{.46\linewidth} 
           \raggedright{(\textit{b})}
		       \includegraphics[width=\linewidth]
           {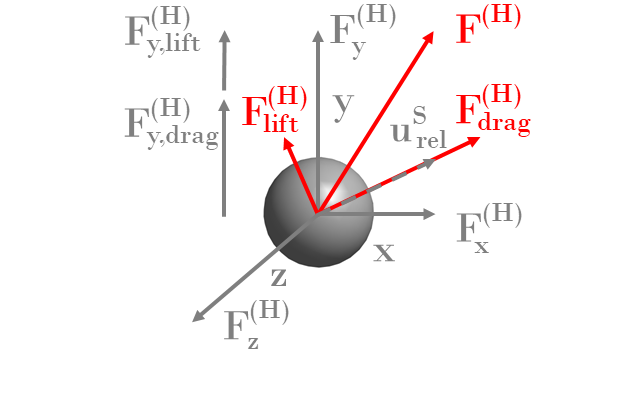}
         \end{minipage}

         \caption{(\textit{a}) A spherical surface with radius $R_s$, trimmed by two wall-parallel planes at one particle radius $R_\mathrm{p}$, over which a fluid velocity is evaluated for the purpose of calculating the instantaneous relative velocity of the fluid in the vicinity of the particle. Black and red points on the spherical surface correspond to the sampling points $\mathbf{x}^{(i)}_l$ used for the fluid velocity, while the red points denote the subset of sampling points used for evaluating the local shear rate around the particle. (\textit{b}) Schematic of the hydrodynamic forces acting on the particle. The relative velocity vector $\mathbf{u}_{\text{rel}}^{\mathrm{S}}$, total hydrodynamic force $\mathbf{F}^\mathrm{(H)}$, and its projections that define the drag and lift forces as well as the drag and lift forces contributions to $\mathbf{F}^\mathrm{(H)}$ in the wall-normal direction.}
    \label{fig:shell_and_forces}
\end{figure}

For the purpose of calculating the relative velocity of the particles, we adopted the approach proposed by \citet{kidanemariam2013direct}. The instantaneous relative velocity of the $i$th particle is given by 

\begin{align}
\mathbf{u}_{\text{rel}}^S(i, R_s, t) 
= \mathbf{u}_\mathrm{f}^S(\mathbf{x}^{(i)}_\mathrm{p}, R_s, t) 
- \mathbf{u}^{(i)}_\mathrm{p}(t),
\end{align}
where $\mathbf{u}_\mathrm{f}^S$ denotes the instantaneous velocity of the fluid in the vicinity of the $i$th particle. In order to approximate $\mathbf{u}_\mathrm{f}^S$, first a spherical surface with radius $R_s$ with its center at the same location as the center of the particle $\mathbf{x}_\mathrm{p}^{(i)}$ is taken. Additionally, the spherical surface is trimmed by two wall-parallel planes at one particle radius $R_\mathrm{p}$ below and above the center as shown in \autoref{fig:shell_and_forces}(\textit{a}) to account for the strong wall-normal gradients near the wall. The fluid velocity evaluated through trilinear interpolation at sampling points $\mathbf{x}^{(i)}_l$ located on the surface $S$, results in $\mathbf{u}_\mathrm{f}^S$, which is defined as,

\begin{align}
\mathbf{u}_\mathrm{f}^S(\mathbf{x}^{(i)}_\mathrm{p}, R_s, t) 
&= \frac{1}{n_S^{(i)}(t)} 
    \sum_{l=1}^{N_l} 
    \phi_\mathrm{f}(\mathbf{x}^{(i)}_l(t), t) 
    \mathbf{u}_\mathrm{f}(\mathbf{x}^{(i)}_l(t), t),
    \quad \forall i = 1, \dots, N_\mathrm{p}, \\
\text{where} \quad
n_S^{(i)}(t) 
&= \sum_{l=1}^{N_l} 
    \phi_\mathrm{f}(\mathbf{x}^{(i)}_l(t), t), 
\quad \forall i = 1, \dots, N_\mathrm{p}.
\end{align}

Here $\phi_\mathrm{f}(\mathbf{x},t)$ is the fluid indicator function that avoids undesired sampling of velocity data inside a neighboring particle. $\mathbf{x}^{(i)}_l(t) = \mathbf{x}^{(i)}_\mathrm{p}(t) + R_{s} \, \mathbf{n}_s$ is the position vector of the point $l$ on the surface $S$, with $\mathbf{n}_s$ being the outward pointing unit vector normal to $S$ and $N_l$ is the number of Lagrangian marker points discretizing $S$. In the present simulations, $R_s = 3R_\mathrm{p}$. The relative velocity computed using $R_s$ determines the direction for the decomposition of hydrodynamic forces, discussed in \S~\ref{sec:force_balance_and_dec}. A sensitivity analysis of the decomposed forces for different values of $R_S$ is presented in \autoref{sec:appendix_rs}.

\subsection{Entrainment threshold}
\label{sec:entrainment_threshold}

For the particle to be lifted off from the plane lower wall surface, the hydrodynamic forces acting on the particle need to exceed the weight of the particle, and in this study, we are interested in which mechanism causes forces to increase beyond that threshold. Occasionally, a particle might exhibit small detachments from the wall before settling back to the wall in a short time period. Therefore, the wall-normal threshold distance $y_{\mathrm{th}}$ is selected to capture particles that do not fall back too soon after they start an upward motion.

Particles are initially located at a wall-normal distance of one particle diameter $D_\mathrm{p}$ from the wall. At the beginning of the simulations, due to the gravitational settling effect, particles quickly settle toward the wall, resulting in the particles' wall-normal coordinate being $y_\mathrm{p} \approx y_\mathrm{w}$, where $ y_\mathrm{w} = R_\mathrm{p} + 2\Delta x$, the small offset being due to the repulsive force discussed in \S~\ref{sec:numerical_method}. The detection criteria for an entrainment event are then chosen as $y_\mathrm{p}$ crossing the $y_{\mathrm{th}}$ threshold line in increasing $y_\mathrm{p}$ direction, where $y_{\mathrm{th}} = D_\mathrm{p} + 2\Delta x$. 

\subsection{Relevant time interval}
\label{sec:relevant_time}

In order to investigate the mechanism of particle lift-off, it is useful to specify the temporal interval over which the lift-off progresses. For this purpose, we define an interval that ends at the time $t_\mathrm{end}$ when a particle crosses the wall-normal coordinate $y_\mathrm{th}$ from below, as described previously. The first observation to be made is particles crossing the entrainment threshold, which has already been discussed in \S~\ref{sec:entrainment_threshold}. The second observation to be made concerns the wall-normal velocity $v_\mathrm{p}$ of particles. When particles start upward motion, their $v_\mathrm{p}$ starts to become positive. Therefore, once the entrainment is detected for the $j$th event, we assign $t_\mathrm{end}^{j}$ to the instant when the $y_\mathrm{th}$ line is crossed. Afterwards, by going backwards in time from $t_\mathrm{end}^{j}$, the start of entrainment $t_\mathrm{start}^{j}$ is designated as the instant of time with the first $v_\mathrm{p}^{j} > 10^{-5}$. The relevant time interval $[t_\mathrm{start}^{j}, t_\mathrm{end}^{j}]$ is therefore apriori different for each entrainment event. To compare all events, the simulation time is shifted to the start of each event, and it is represented as $t_\mathrm{r} = t_\mathrm{obs} - t_\mathrm{start}^{j}$  for each $j \in N_\mathrm{e}$, where $t_\mathrm{obs}$ denotes the original simulation time and $N_\mathrm{e}$ is the number of entrainment events. As a result, all events are aligned such that $t_\mathrm{r} = 0$ corresponds to the beginning of each event.

\subsection{Force balance and decomposition}
\label{sec:force_balance_and_dec}

\begin{figure}[t]
\raggedleft
    \begin{tikzpicture}
        \node[anchor=south west,inner sep=0] (fig) at (0,0) 
            {\includegraphics[width=0.94\linewidth]{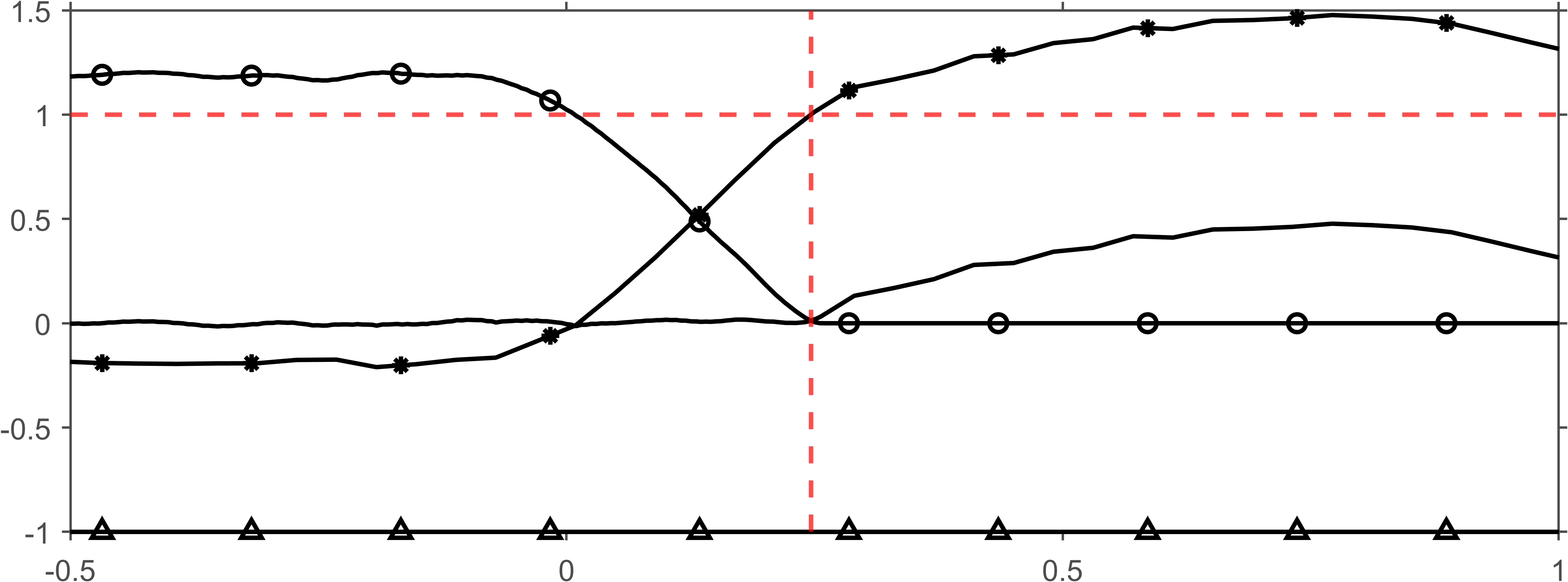}};
        \node at ($(fig.south west)!0.5!(fig.south east)$) [below=3mm, anchor=center, xshift=3.5mm] {$t_{\mathrm{r}}/t_{\mathrm{b}}$};
        \node[rotate=90, anchor=center, yshift=3.5mm, xshift=2.5mm] at ($(fig.south west)!0.5!(fig.north west)$) {$F_y/F_g$};
        \node[anchor=north west, xshift=-7mm, yshift=5mm] at (fig.north west) {(\textit{a})};
    \end{tikzpicture}

    \begin{tikzpicture}
        \node[anchor=south west,inner sep=0] (fig) at (0,0) 
            {\includegraphics[width=0.94\linewidth]{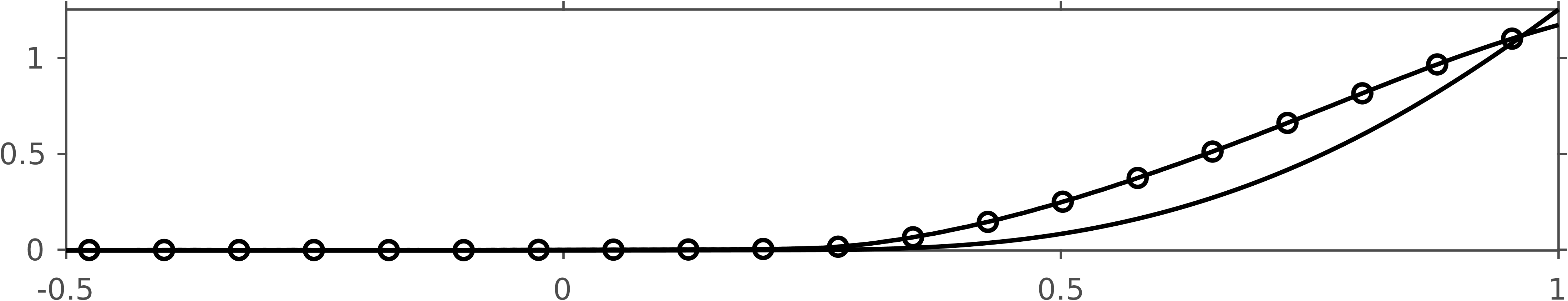}};
        \node at ($(fig.south west)!0.5!(fig.south east)$) [below=3mm, anchor=center, xshift=3.5mm] {$t_{\mathrm{r}}/t_{\mathrm{b}}$};
        \node[rotate=90, anchor=center, yshift=3.5mm, xshift=2.5mm] at ($(fig.south west)!0.5!(fig.north west)$) {$v_{\mathrm{p}}/u_{\tau}$, $y_{\mathrm{p,w}}/R_{\mathrm{p}}$};
        \node[anchor=north west, xshift=-7mm, yshift=7mm] at (fig.north west) {(\textit{b})};
    \end{tikzpicture}
    \caption{(\textit{a}) Time evolution of wall-normal hydrodynamic force $F_y^{(\mathrm{H})}$ (\textasteriskcentered), contact force $F_{\mathrm{c}}$  (\(\circ\)), particle submerged weight $F_g$ (\(\scalebox{0.7}{$\triangle$}\)), and wall-normal total force $F_y$ (-) of one entrained particle. Red dashed lines indicate $F_y^{(\mathrm{H})}=-F_g$ (horizontal) and $F_{\mathrm{c}} = 0$ (vertical). Their intersection marks the time when both are satisfied. (\textit{b}) Time evolution of wall-normal velocity of the particle $v_{\mathrm{p}}$ (\(\circ\)) and wall-normal position of the particle center $y_{\mathrm{p,w}}$ (-) of one entrained particle. Here $y_\mathrm{p,w} = y_\mathrm{p} - y_\mathrm{w}$. Note that this data corresponds to the same entrainment event as in (\textit{a}).}
    \label{fig:yp_vp_force}
\end{figure}
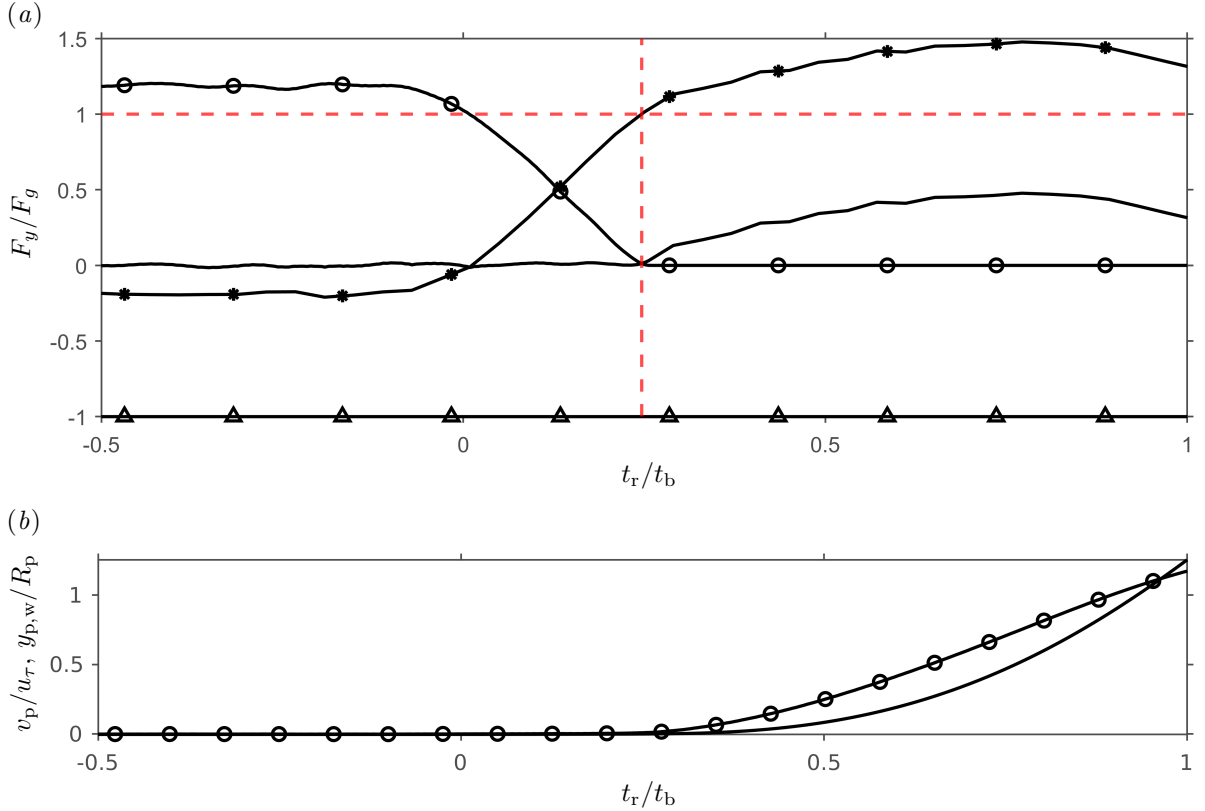

Hydrodynamic forces acting on a particle in streamwise, wall-normal, and spanwise directions are obtained from PR-DNS. The total force $\mathbf{F}=(F_x, F_y, F_z)$ exerted on the particle consists of submerged weight $F_g$ directed along the negative wall-normal axis, contact force $\mathbf{F}_\mathrm{c}$ described in \S~\ref{sec:numerical_method}, 
and hydrodynamic forces $\mathbf{F}^{(\mathrm{H})}=(F_x^{(\mathrm{H})},F_y^{(\mathrm{H})},F_z^{(\mathrm{H})})$. Therefore, \eqref{eq:particle-momentum} can be written as 

\begin{align}
    \rho_\mathrm{p} V_\mathrm{p} \frac{d \mathbf{u}_\mathrm{p}}{dt} &= \mathbf{F}^{(\mathrm{H})} + \mathbf{F}_\mathrm{c} + \mathbf{F}_g, 
    \label{eq:particle-momentum-forces}
\end{align}
where the particle superscript has been omitted for clarity. In \eqref{eq:particle-momentum-forces} $\mathbf{F}^{(\mathrm{H})}=\rho_\mathrm{f} \oint_{\partial I} \boldsymbol{\tau} \cdot \mathbf{n}_I \, d\sigma$, $\mathbf{F}_g = (0,-F_g,0)$ with $F_g=(\rho_\mathrm{p} - \rho_\mathrm{f}) V_\mathrm{p} \mathbf{g}$, and $\mathbf{F}_\mathrm{c}$ is the force due to the particle-particle and particle-wall contact \citep{glowinski1999distributed}. Therefore, the total force $\mathbf{F}$ is given by $\mathbf{F} = \mathbf{F}^{(\mathrm{H})} + \mathbf{F}_\mathrm{c} + \mathbf{F}_g$. To illustrate the relative contributions of these forces, \autoref{fig:yp_vp_force}(\textit{a}) shows the wall-normal component $F_y$ of the total force and its components for a representative entrainment event. As observed, the contact force is active only when the particle is in the immediate vicinity of the wall (cf. \autoref{fig:yp_vp_force}\textit{b}), meaning $y_\mathrm{p} \leq y_\mathrm{w}$, and it decreases to zero as $v_\mathrm{p}$ starts increasing and the wall-normal distance between the particle center and the wall exceeds $y_\mathrm{w}$, which happens at $t_\mathrm{r}/t_\mathrm{b} = 0.25$ in this example. 

In turbulent flows, the hydrodynamic force components in the streamwise $F_x^{(\mathrm{H})}$ and wall-normal directions $F_y^{(\mathrm{H})}$ do not necessarily correspond directly to the classical drag and lift force definitions, respectively, due to the chaotic nature of turbulence. For instance, during a sweep event, the downward motion of the fluid can cause the drag force to deviate from the streamwise direction and align with the direction of the local flow structure. For that reason, we use a projection upon the relative velocity $\mathbf{u}_{\text{rel}}^{\mathrm{S}}$ (cf. \S~\ref{sec:how_to_relvel}).

Hence, the drag force is defined as
\begin{equation}
\mathbf{F}_{\text{drag}}^{(\mathrm{H})} = (\mathbf{F}^{(\mathrm{H})} \cdot \hat{\mathbf{u}}_{\text{rel}}^S)\, \hat{\mathbf{u}}_{\text{rel}}^S,
\quad\text{where}\quad
\hat{\mathbf{u}}_{\text{rel}}^S = \frac{\mathbf{u}_{\text{rel}}^S}{\|\mathbf{u}_{\text{rel}}^S\|},
\end{equation}
and the force orthogonal to the projected force becomes the lift force 
\begin{align}
\mathbf{F}_{\text{lift}}^{(\mathrm{H})} &= \mathbf{F}^{(\mathrm{H})} - \mathbf{F}_{\text{drag}}^{(\mathrm{H})}.
\end{align}
The contributions of $\mathbf{F_{\text{lift}}}^{(\mathrm{H})}$ and $\mathbf{F}_{\text{drag}}^{(\mathrm{H})}$ to $F_y^{(\mathrm{H})}$ are
\begin{equation}
F_{y,\text{drag}}^{(\mathrm{H})} = \mathbf{F}_{\text{drag}}^{(\mathrm{H})} \cdot \hat{\mathbf{e}}_y,
\quad
F_{y,\text{lift}}^{(\mathrm{H})} = \mathbf{F}_{\text{lift}}^{(\mathrm{H})} \cdot \hat{\mathbf{e}}_y,
\quad\text{where}\quad
\hat{\mathbf{e}}_y = (0,\, 1,\, 0),
\end{equation}
and the total hydrodynamic force in the wall-normal direction is
\begin{equation}
F_y^{(\mathrm{H})} = \mathbf{F}^{(\mathrm{H})} \cdot \hat{\mathbf{e}}_y = F_{y,\text{drag}}^{(\mathrm{H})} + F_{y,\text{lift}}^{(\mathrm{H})}.
\end{equation}
The schematic of hydrodynamic forces and force decomposition is represented in \autoref{fig:shell_and_forces}(\textit{b}).

$F_y^{(\mathrm{H})}$ can also be represented as the sum of the forces due to pressure and viscous stresses. The hydrodynamic force $\mathbf{F}^{(\mathrm{H})}$ exerted on the particle is defined as
\begin{align}
\mathbf{F}^{(\mathrm{H})} = -\int_I p \, \mathbf{n_I} \, dI 
+ \int_I \boldsymbol{\tau}_\mathrm{v}\cdot \mathbf{n_I} \, dI,
\label{eq:presviscintegral}
\end{align}

where the viscous stress tensor is given by $\boldsymbol{\tau}_\mathrm{v} = \rho_\mathrm{f} \nu_\mathrm{f} \left( \nabla \mathbf{u}_\mathrm{f} + (\nabla \mathbf{u}_\mathrm{f})^\top \right)$. Consequently, to investigate pressure and viscous stress contributions to the total hydrodynamic $F_y^{(\mathrm{H})}$ force, we take 
\begin{equation}
F_{y,\text{pres}}^{(\mathrm{H})} = \mathbf{F}_{\text{pres}}^{(\mathrm{H})} \cdot \hat{\mathbf{e}}_y,
\quad
F_{y,\text{visc}}^{(\mathrm{H})} = \mathbf{F}_{\text{visc}}^{(\mathrm{H})} \cdot \hat{\mathbf{e}}_y,
\end{equation}
where $\mathbf{F}_{\text{pres}}^{(\mathrm{H})}$ is the first term of the right-hand side of \eqref{eq:presviscintegral} and $\mathbf{F}_{\text{visc}}^{(\mathrm{H})}$ is the second term.

\subsection{The action of vortices}
\label{sec:action_of_cortices}

\begin{figure}
\raggedleft
    \begin{tikzpicture}
        \node[anchor=south west,inner sep=0] (fig) at (0,0) 
            {\includegraphics[width=0.95\linewidth]{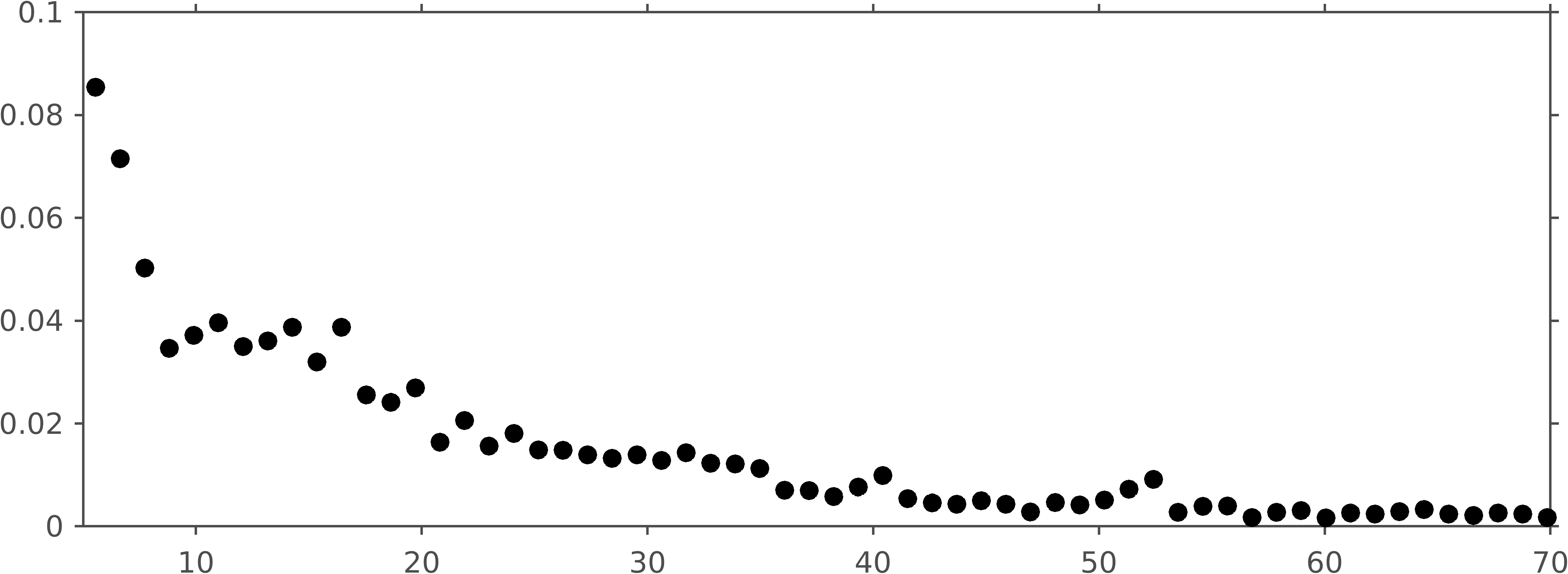}};
        \node at ($(fig.south west)!0.5!(fig.south east)$) [below=3mm, anchor=center, xshift=3.5mm] {$V_{\boldsymbol{\omega}_{\mathrm{f}}} / V_{\mathrm{p}}$};
        \node[rotate=90, anchor=center, yshift=3.5mm, xshift=2.5mm] at ($(fig.south west)!0.5!(fig.north west)$) {P.d.f.};
        \node[anchor=south west,inner sep=0] (inset) at ($(fig.north east)+(-10cm,-4cm)$) 
        {\includegraphics[width=0.6\linewidth]{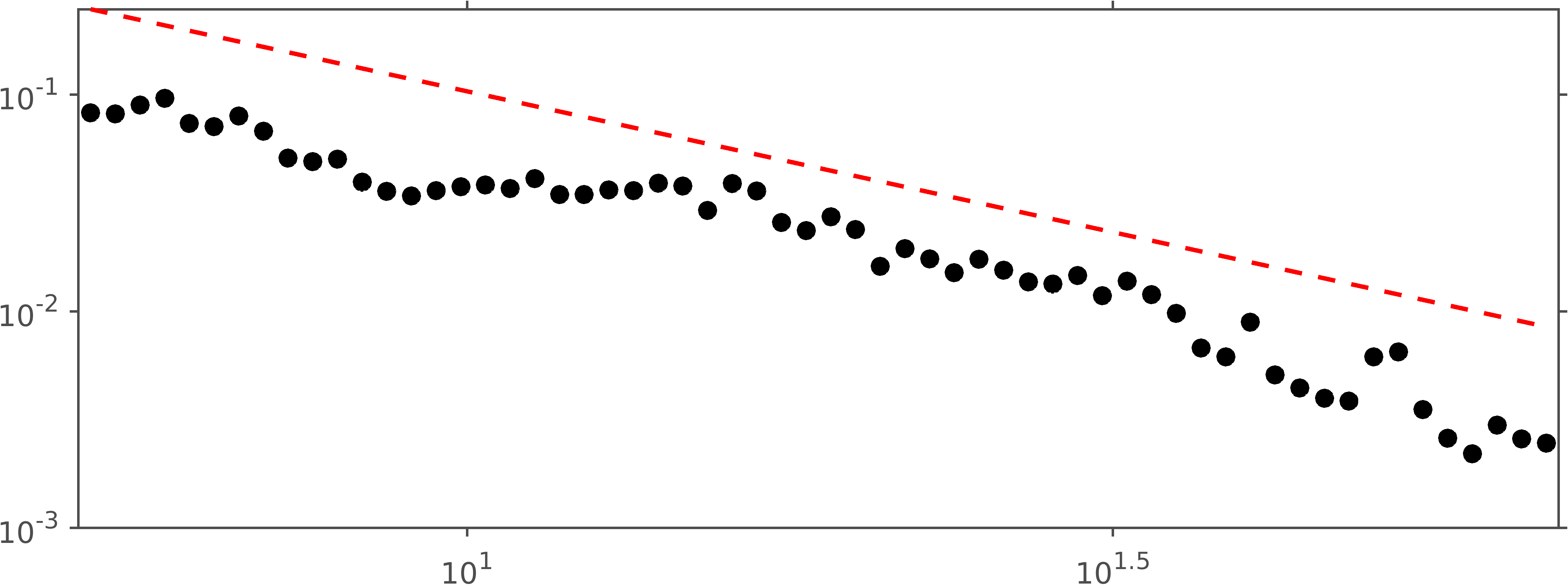}};
    \end{tikzpicture}
    \caption{The P.d.f. of the vortex volumes normalized by particle volume for the unconstrained case. The inset shows the same data on a log--log scale, highlighting power-law behavior ($V_{\boldsymbol{\omega}_{\mathrm{f}}}^{(-1.3)}$).}
    \label{fig:vortices_vol}
\end{figure}

Coherent structures have been identified as playing a crucial role in the dynamics of wall-bounded turbulent flows \citep{annurev:/content/journals/10.1146/annurev.fl.23.010191.003125,jimenez2018coherent}. As mentioned in the introduction, researchers have provided evidence of the presence of intense vortical structures in the vicinity of a particle during entrainment in an open channel flow. In order to explore the proximity of vortices during particle entrainment, we determine the distance from each particle to the nearest vortex. In this study, the $Q$ criterion \citep{hunt1988eddies} is employed to identify the vortex structures. The vortical regions are found using the second invariant of $\nabla \mathbf{u}_\mathrm{f}$ which for incompressible flows is given by $Q = \frac{1}{2}\left( \| \boldsymbol{\Omega} \|^2 - \| \mathbf{S} \|^2 \right)$ where $\boldsymbol{\Omega}$ is the rotation tensor and $\mathbf{S}$ is the rate-of-strain tensor. Then the criterion for finding the outline of a vortex is set to $Q = Q_\mathrm{th}$ with $Q_\mathrm{th}$ being some positive threshold. In the present study $Q_\mathrm{th} = 4 Q_\mathrm{rms}$, where $Q_\mathrm{rms}$ is the root-mean-square of the $Q$ field and $Q^+_\mathrm{rms} = 0.02$. We calculate the vortex volume $V_{\boldsymbol{\omega}_{\mathrm{f}}}$ from a triangulation of the surface identified by the $Q$ criterion and a surface integral (using the divergence theorem). The distance from the particle center to the closest point on any vortex surface is denoted by $d_{\boldsymbol{\omega}_{\mathrm{f}}}$. Small vortical structures attached to or located near the solid particles form frequently. In order to eliminate their signature on the distance statistics, we impose a threshold on $V_{\boldsymbol{\omega}_{\mathrm{f}}}$ and disregard vortices that have an enclosed volume $V_{\boldsymbol{\omega}_{\mathrm{f}}}$ smaller than 4.5$V_\mathrm{p}$ before calculating $d_{\boldsymbol{\omega}_{\mathrm{f}}}$. Note that for computational convenience, the vortex eduction has only been carried out for the lower half of the computational domain $y = [0, 0.5h]$. This method has previously been used by \citet{uhlmann2017clustering}. \autoref{fig:vortices_vol} represents the resulting probability density of $V_{\boldsymbol{\omega}_{\mathrm{f}}}$. We observe an approximate power-law scaling for vortex volumes with $V_{\boldsymbol{\omega}_{\mathrm{f}}}^{(-1.3)}$.

\section{Results}
\label{sec:result}
\subsection{Global lift-off statistics}
\label{sec:global_liftoff_stats} 

The frequencies of detected entrainment events in different simulations are $fr^{(\text{xx})} = N_{\mathrm{e}}^{(\text{xx})^*} / (t^{(\text{xx})}_\mathrm{obs}/t_\mathrm{b})$, where xx stands for the specific series of simulations, $N_{\mathrm{e}}^{(\text{xx})^*}$ is the total number of detected entrainment events, and $t^{(\text{xx})}_\mathrm{obs}$ is the observation time of the simulations. The corresponding values are listed in \autoref{tab:frequencies_table}. The total number of entrainment events is $ N_{\mathrm{e}}^{(\mathrm{UN})^*} = 68$, $ N_{\mathrm{e}}^{(\mathrm{RC})^*} = 18$, and $ N_{\mathrm{e}}^{(\mathrm{ZC})^*} = 18$. When rotation was suppressed in the simulations, no significant difference in the frequency of entrainment events was detected, suggesting that the rotation of particles does not play an important role in the entrainment process. In turn, suppressing spanwise motion for particles increased the frequency significantly, possibly already implying that the particles sampling more flow structures has an effect on the entrainment. The latter is further supported by the observed average particle streamwise velocities for different simulations (cf. \autoref{tab:frequencies_table}). The averaging is performed on the particles that did not lift off. $\langle u_\mathrm{p}^{\mathrm{ZC}} \rangle$ has higher values compared to $\langle u_\mathrm{p}^{\mathrm{UN}} \rangle$ and $\langle u_\mathrm{p}^{\mathrm{RC}} \rangle$, implying that in the ZC case, particles sample flow structures other than low-speed regions more frequently than in the UN and RC cases.

Particle resuspension can occur due to several physical mechanisms, including the collision-induced particle-particle and particle-wall resuspension \citep{HENRY20231}. It is essential to note that this study focuses on the mechanism behind the entrainment of particles caused by flow structures. Hence, after detecting events that satisfy the entrainment threshold (cf. \S~\ref{sec:entrainment_threshold}), we exclude those that feature particle-particle and particle-wall collisions before $t_{\mathrm{start}}$, within a time interval that could have led to their entrainment, as well as events featuring particles that collide with another particle during the entrainment period defined in \S~\ref{sec:relevant_time}. Here, particle-wall collision specifically refers to events in which a particle is temporarily detached, returns, and hits the wall at a finite relative velocity. Note that if the particles are not lifted from the wall, they remain in continuous contact with it. The filtering results in $ N_{\mathrm{e}}^{(\mathrm{UN})} = 20$, $N_{\mathrm{e}}^{(\mathrm{RC})} = 8$, and $N_{\mathrm{e}}^{(\mathrm{ZC})} = 8$ analyzed events. Moreover, the number of entrainment events in the present study (20 for the unconstrained case), although small, is comparable to that reported by \citet{cameron2020entrainment}, who analyzed 25 events.

\begin{table}[t]
\raggedleft
\caption{Entrainment statistics in different simulations. $N_{\mathrm{e}}^*$ denotes total number of detected entrainment events, $t_{\mathrm{obs}}$ is observation time, $fr$ is frequency of entrainment events with respect to $N_{\mathrm{e}}^*$, $N_{\mathrm{e}}$ represents number of analyzed entrainment events, $\langle t_{\mathrm{rel}} \rangle$ is average relevant time, $\langle u_\mathrm{p} \rangle$ is the streamwise particle velocity, $\langle F_y^{(\mathrm{H})} \rangle$ denotes total hydrodynamic force in wall-normal direction, $\langle S_{\mathrm{g}} \rangle$ is global wall-normal shear around the particles, and $\langle d_{\boldsymbol{\omega}_{\mathrm{f}}\mathrm{g}} \rangle$ is global distance to the nearest vortical structure.}
\label{tab:frequencies_table}
\begin{tabular}[t]{lccc}
\hline
&Unconstrained (UN)&Rotation-constrained (RC)&Spanwise-constrained (ZC)\\
\hline
$N_{\mathrm{e}}^*$&68&18&18\\
$t_{\mathrm{obs}}/t_\mathrm{b}$&282&55&26\\
$fr \cdot t_\mathrm{b}$&0.36&0.48&1.03\\
$N_{\mathrm{e}}$&20&8&8\\
$\langle t_{\mathrm{rel}} \rangle /t_{\mathrm{b}}$&0.91&0.99&1.44\\
$\langle u_\mathrm{p} \rangle / u_\tau$&3.27&3.35&4.69\\
$\langle F_y^{(\mathrm{H})} \rangle / F_g$&0.25&0.26&0.21\\
$\langle S_{\mathrm{g}} \rangle D_\mathrm{p}/u_\tau$&4.81&4.34&5.59\\
$\langle d_{\boldsymbol{\omega}_{\mathrm{f}}\mathrm{g}} \rangle/D_\mathrm{p}$&15.55&16.31&16.41\\
\hline
\end{tabular}
\end{table}

The average time a particle takes to cross the $y_{\mathrm{th}}$ line from $t_{\mathrm{start}}$ is denoted as $\langle t_{\mathrm{rel}}^{(\mathrm{xx})} \rangle / t_{\mathrm{b}}$, where $t_{\mathrm{rel}} = t_{\mathrm{end}} - t_{\mathrm{start}}$, and is given in \autoref{tab:frequencies_table}. The relative time interval has been scaled using the bulk fluid time scale $t_\mathrm{b}$, which represents the characteristic time of large-scale flow motions. Since in this study, we are interested in particle entrainment in relation to flow structures, using the bulk time scale allows for a consistent comparison of entrainment dynamics to large turbulent structures motion. Furthermore, as can be seen in \autoref{tab:frequencies_table}, the relevant time interval $t_\mathrm{rel}$ is of order unity when scaled with bulk fluid time. We observe that the average duration of entrainment events $\langle t_\mathrm{rel} \rangle$ in the spanwise constrained case (ZC) is significantly larger than for unconstrained particles.

\subsection{Force decomposition}
\label{sec:focrs_dec}

Before discussing the components of lift and drag to the vertical hydrodynamic force, we first consider the force balance for a particle during an entrainment event in more detail. \autoref{fig:yp_vp_force}(\textit{a}) presents the force balance prior to the onset of entrainment, that is, before $t_\mathrm{r} = 0$, and during the relevant time interval. Before time $t_\mathrm{r} = 0$, the particle remains in contact with the wall. During this period, the hydrodynamic force $F_y^{(\mathrm{H})}$ acts in the negative wall-normal direction, pushing the particle toward the wall, while the contact force $F_\mathrm{c}$ counterbalances both $F_y^{(\mathrm{H})}$ and the submerged particle weight $F_g$. As $F_y^{(\mathrm{H})}$ increases, the particle begins to move very slightly in the positive wall-normal direction, such that $F_\mathrm{c}$ gradually decreases. The total wall-normal force $F_y$ remains zero, as $F_y^{(\mathrm{H})}$ must first exceed $F_g$ before the particle can detach. In this particular case, this occurs at approximately $t_\mathrm{r}/t_\mathrm{b} = 0.25$, when the particle is no longer in contact with the wall. Following this instant, changes in the particle wall-normal velocity $v_\mathrm{p}$ and wall-normal coordinate $y_\mathrm{p,w}$ become apparent (cf. \autoref{fig:yp_vp_force}\textit{b}). In this specific event, the particle crosses the threshold wall distance at $t_\mathrm{r}/t_\mathrm{b} = 1.26$, which lies beyond the temporal range shown in the figure.

\begin{figure}[t]
\raggedleft
    \begin{tikzpicture}
        \node[anchor=south west,inner sep=0] (fig) at (0,0) 
            {\includegraphics[width=0.95\linewidth]{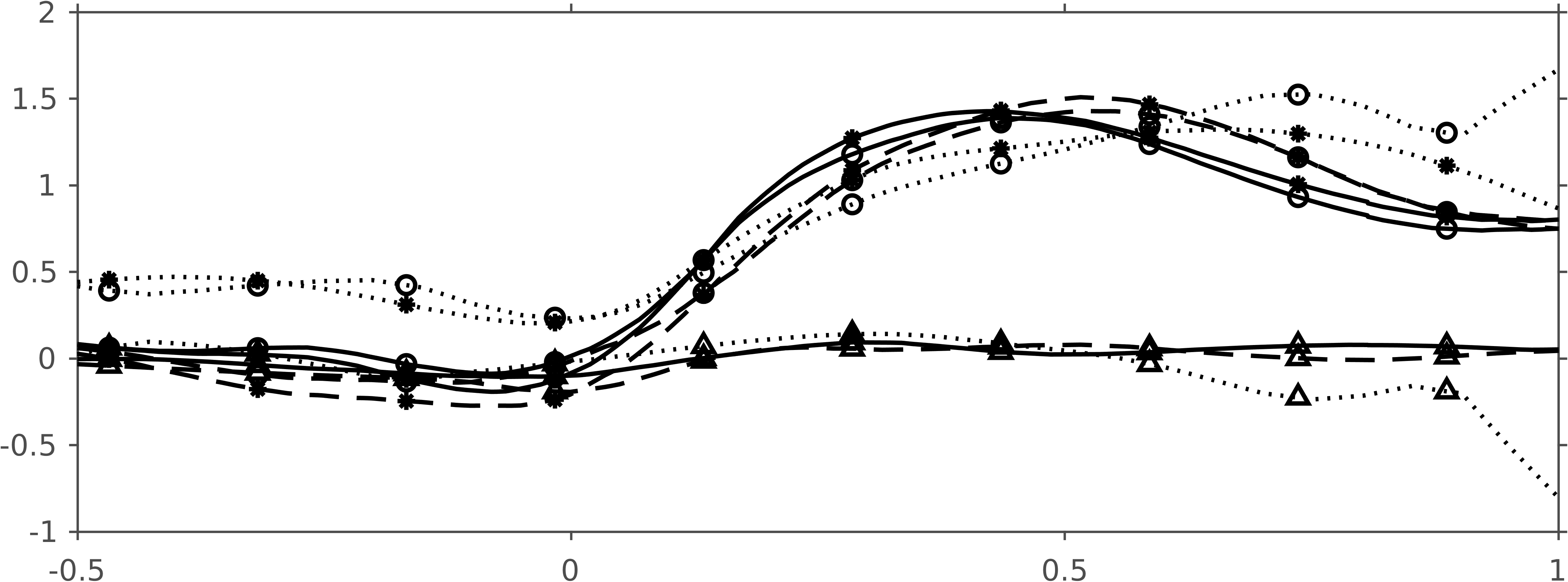}};
        \node at ($(fig.south west)!0.5!(fig.south east)$) [below=3mm, anchor=center, xshift=3.5mm] {$t_{\mathrm{r}}/t_{\mathrm{b}}$};
        \node[rotate=90, anchor=center, yshift=3.5mm, xshift=2.5mm] at ($(fig.south west)!0.5!(fig.north west)$) {$\langle F_y^{(\mathrm{H})} \rangle /F_g$};
    \end{tikzpicture}
    \caption{Time evolution of hydrodynamic force in wall-normal direction $F_y^{(\mathrm{H})}$ (\textasteriskcentered) with lift $F_{y,\mathrm{lift}}^{(\mathrm{H})}$ (\(\circ\)) and drag $F_{y,\mathrm{drag}}^{(\mathrm{H})}$ (\(\scalebox{0.7}{$\triangle$}\)) contributions, The data is averaged over all corresponding entrainment events for UN (solid line), RC (dashed line), and ZC (dotted line) cases. }
    \label{fig:lift_drag_avg}
\end{figure}

The two components of the total wall-normal hydrodynamic force and $F_y^{(\mathrm{H})}$, averaged over the detected entrainment events, are shown in \autoref{fig:lift_drag_avg}. The forces are normalized by the reference force $F_g$ throughout the study. The lift $F_{y,\mathrm{lift}}^{(\mathrm{H})}$ and drag $F_{y,\mathrm{drag}}^{(\mathrm{H})}$ contributions to $F_y^{(\mathrm{H})}$, as well as other decompositions, are obtained as described in \S~\ref{sec:force_balance_and_dec}. Before $t_\mathrm{r} = 0$, the total hydrodynamic force remains small (even negative) for cases UN, RC, and then increases as the particle detaches and overcomes its submerged weight. For the spanwise-constrained case (ZC), we observe that the hydrodynamic force already features a significant positive value before $t_\mathrm{r}$, despite the fact that the global (unconstrained) average value for this case is similar to the corresponding values for cases UN, RC, cf. \autoref{tab:frequencies_table}. Note that $F_y^{(\mathrm{H})}/F_g=1$ is on average reached after roughly 0.25$t_\mathrm{b}$ has elapsed. The time evolution data for individual entrainment events are similar to these average curves. The dominant contribution to the total force arises from the lift component $F_{y,\mathrm{lift}}^{(\mathrm{H})}$. Although a small increase in $F_{y,\mathrm{drag}}^{(\mathrm{H})}$ is observed around $t_\mathrm{r} = 0 $, it is insufficient to induce particle entrainment. The observed average increase of the lift component around $t_\mathrm{r}=0$ is common to all three cases, implying that hydrodynamic lift plays the dominant role in initiating particle entrainment. 

\begin{figure}
\raggedleft
    \begin{tikzpicture}
        \node[anchor=south west,inner sep=0] (fig) at (0,0) 
            {\includegraphics[width=0.95\linewidth]{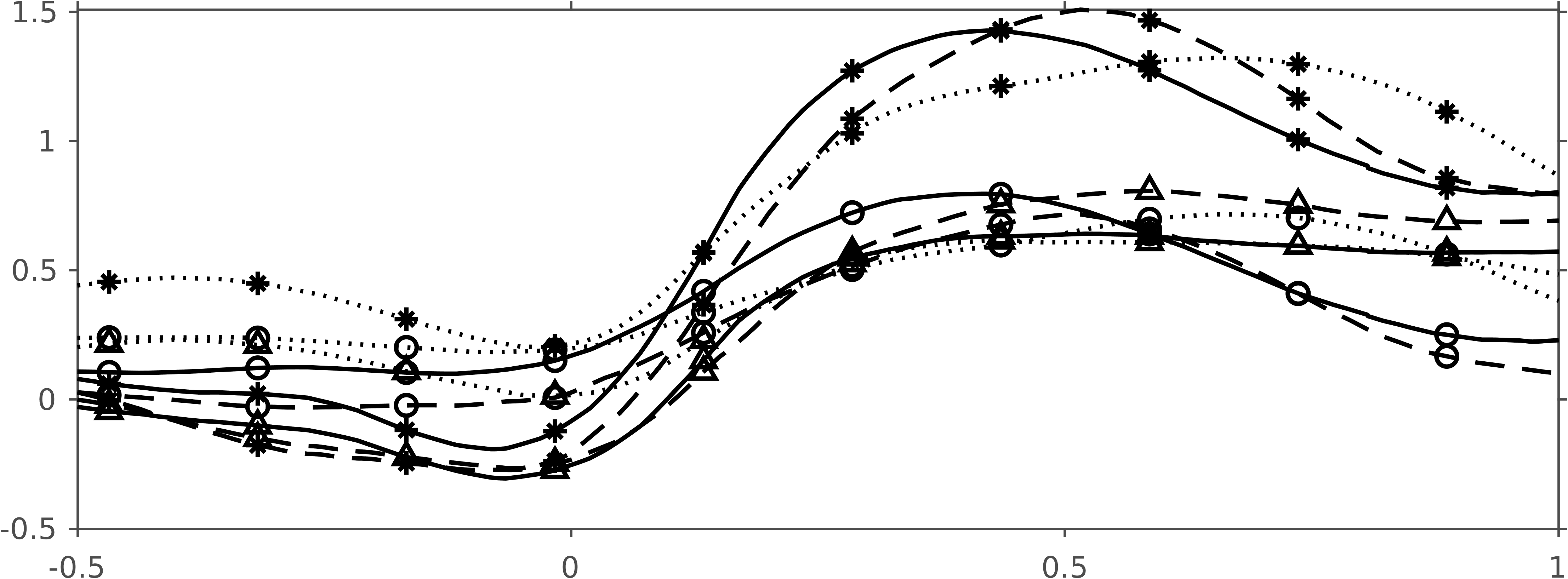}};
        \node at ($(fig.south west)!0.5!(fig.south east)$) [below=3mm, anchor=center, xshift=3.5mm] {$t_{\mathrm{r}}/t_{\mathrm{b}}$};
        \node[rotate=90, anchor=center, yshift=3.5mm, xshift=2.5mm] at ($(fig.south west)!0.5!(fig.north west)$) {$\langle F_y^{(\mathrm{H})} \rangle /F_g$};
    \end{tikzpicture}
    \caption{Time evolution of wall-normal hydrodynamic force $F_y^{(\mathrm{H})}$ (\textasteriskcentered) with pressure $F_{y,\mathrm{pres}}^{(\mathrm{H})}$ (\(\circ\)) and viscous $F_{y,\mathrm{visc}}^{(\mathrm{H})}$ (\(\scalebox{0.7}{$\triangle$}\)) contributions averaged over all corresponding entrainment events for UN (solid line), RC (dashed line), and ZC (dotted line) cases. }
    \label{fig:pres_visc_avg}
\end{figure}

\autoref{fig:pres_visc_avg} shows the pressure contribution $F_{y,\mathrm{pres}}^{(\mathrm{H})}$ and viscous stresses contribution $F_{y,\mathrm{visc}}^{(\mathrm{H})}$ to $F_y^{(\mathrm{H})}$, averaged over the subset of respective events. $F_{y,\mathrm{pres}}^{(\mathrm{H})}$ and $F_{y,\mathrm{visc}}^{(\mathrm{H})}$ are observed to increase after the particle upward motion starts, and it is clearly seen that both forces contribute to the particle lift-off in almost equal proportions. This is consistent with previous experimental studies, which showed how flow structures induce differential pressure acting on sediment particles on a rough substrate \citep{dwivedipressure,dwivedi2011flow}. The trends for UN and RC cases observed in \autoref{fig:lift_drag_avg} and \autoref{fig:pres_visc_avg} further strengthen the discussion about rotation-constrained simulations given in \S~\ref{sec:global_liftoff_stats}, and together with the entrainment frequencies given in \autoref{tab:frequencies_table} indicate that particle rotation can be considered to have a small effect on the entrainment of the particle. 

\subsection{Shear rate}
\label{sec:shear_rate}

\begin{figure}[p]
\raggedleft
    \begin{tikzpicture}
        \node[anchor=south west,inner sep=0] (fig) at (0,0) 
            {\includegraphics[width=0.95\linewidth]{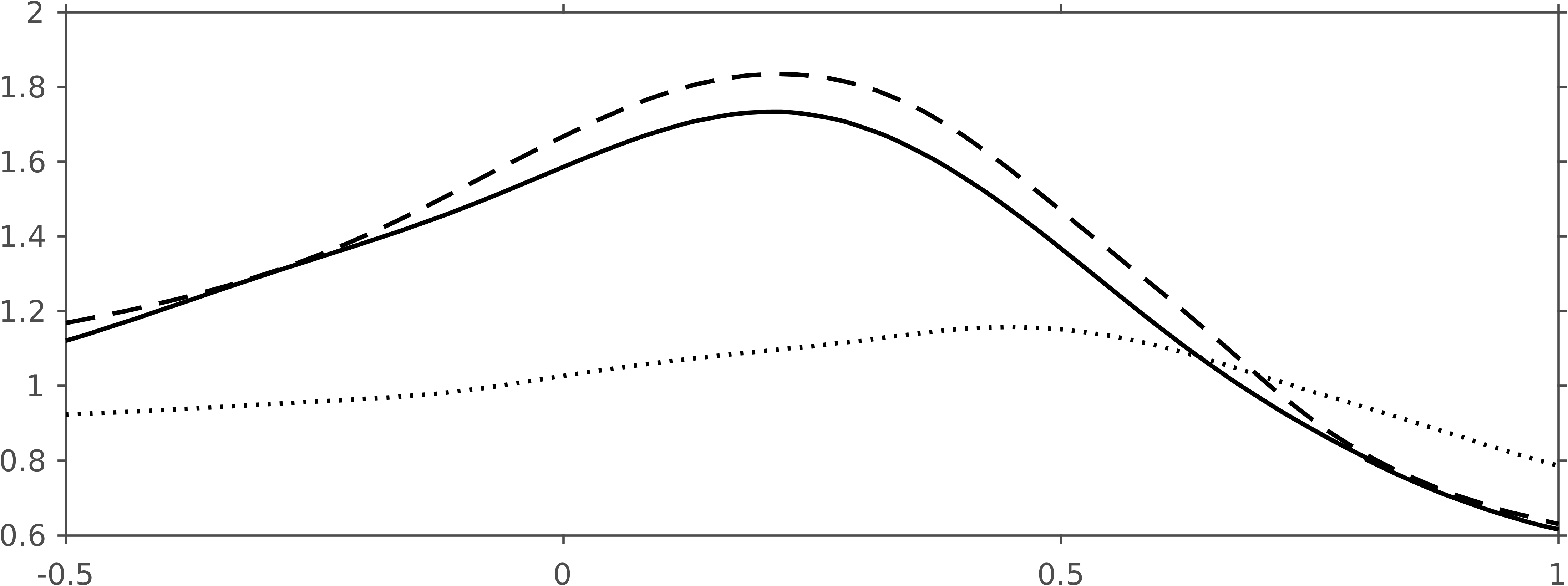}};
        \node at ($(fig.south west)!0.5!(fig.south east)$) [below=3mm, anchor=center, xshift=3.5mm] {$t_{\mathrm{r}}/t_\mathrm{b}$};
        \node[rotate=90, anchor=center, yshift=3.5mm, xshift=2.5mm] at ($(fig.south west)!0.5!(fig.north west)$) {$\langle S_{\mathrm{l}} \rangle / \langle S_{\mathrm{g}} \rangle$};
    \end{tikzpicture}
    \caption{Time evolution of local shear rate $S_{\mathrm{l}}$ for entrained particles averaged over all corresponding entrainment events for UN (solid line), RC (dashed line), and ZC (dotted line) cases. }
    \label{fig:shear_rate}
\end{figure}

The preceding analysis indicates that the lift on the present particles is not spin-induced, since switching off particle rotation did not significantly affect the statistics or the observed sequence of events. To further identify the mechanism responsible for generating lift, \autoref{fig:shear_rate} presents the local shear rate $S_{\mathrm{l}}$ around the particle, averaged over the corresponding entrainment events and normalized by the global shear rate $\langle S_{\mathrm{g}} \rangle$ (cf. \autoref{tab:frequencies_table}). The latter are computed as an average over the non-entrained particles. Local shear around the particle was evaluated at the subset of sampling points $\mathbf{x}^{(i)}_l$ of the spherical surface described in \S~\ref{sec:how_to_relvel} and shown in \autoref{fig:shell_and_forces}(\textit{a}). $S_{\mathrm{l}}$ is observed to increase well before the start of the relative time interval. The peaks of $S_{\mathrm{l}}^{\mathrm{UN}}$ and $S_{\mathrm{l}}^{\mathrm{RC}}$ occur roughly at the time corresponding to 0.25$t_\mathrm{b}$. They exhibit comparable magnitudes and temporal evolution throughout the considered time interval. However, $S_{\mathrm{l}}^{\mathrm{ZC}}$ takes more time to grow beyond the corresponding global average value and reach its peak. It can be expected that an increased wall-normal shear rate will lead to correspondingly increased values of the wall-normal hydrodynamic force acting on the particle under observation \citep{loth2008lift}. Therefore, we conclude that the initiation of the generation of a positive wall-normal force results from high shear acting on the particle in all three simulation series. 

\begin{figure}

        {
        \centering
        \begin{minipage}{0.5\linewidth}
            \includegraphics[width=\linewidth]{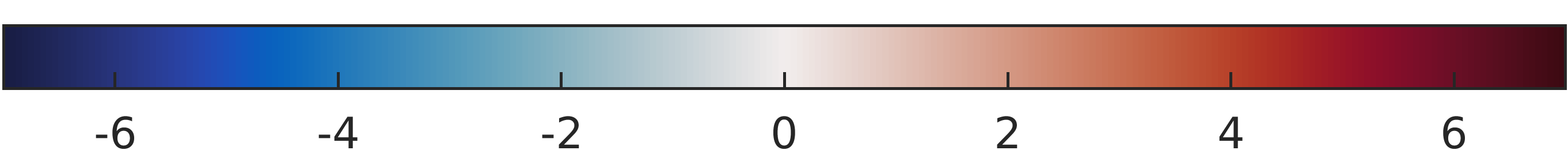}
             \centerline{\small $u_{\mathrm{f}}'/u_\tau$}
        \end{minipage}
        \par
        }
\raggedleft
        \par\vspace{1ex}
         \begin{minipage}{2ex}
           \rotatebox{90}
           {\small \hspace{5ex} $y/h$}
         \end{minipage}
         \begin{minipage}{.46\linewidth} 
		       \raggedright{(\textit{a})}
           \includegraphics[width=\linewidth]
           {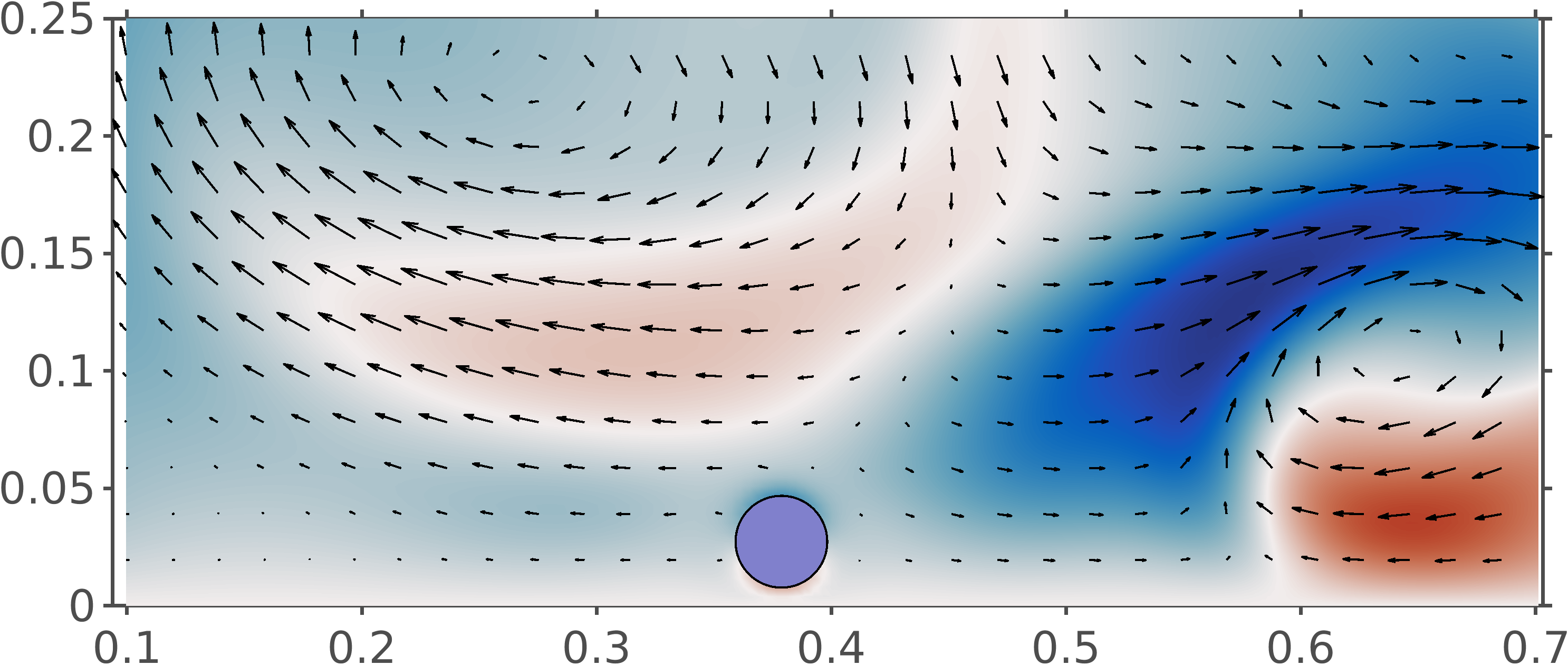}
         \end{minipage}
         \hspace{1ex} 
         \begin{minipage}{2ex}
           \rotatebox{90}
           {\small \hspace{5ex}}
         \end{minipage}
         \begin{minipage}{.46\linewidth} 
           \raggedright{(\textit{b})}
		       \includegraphics[width=\linewidth]
           {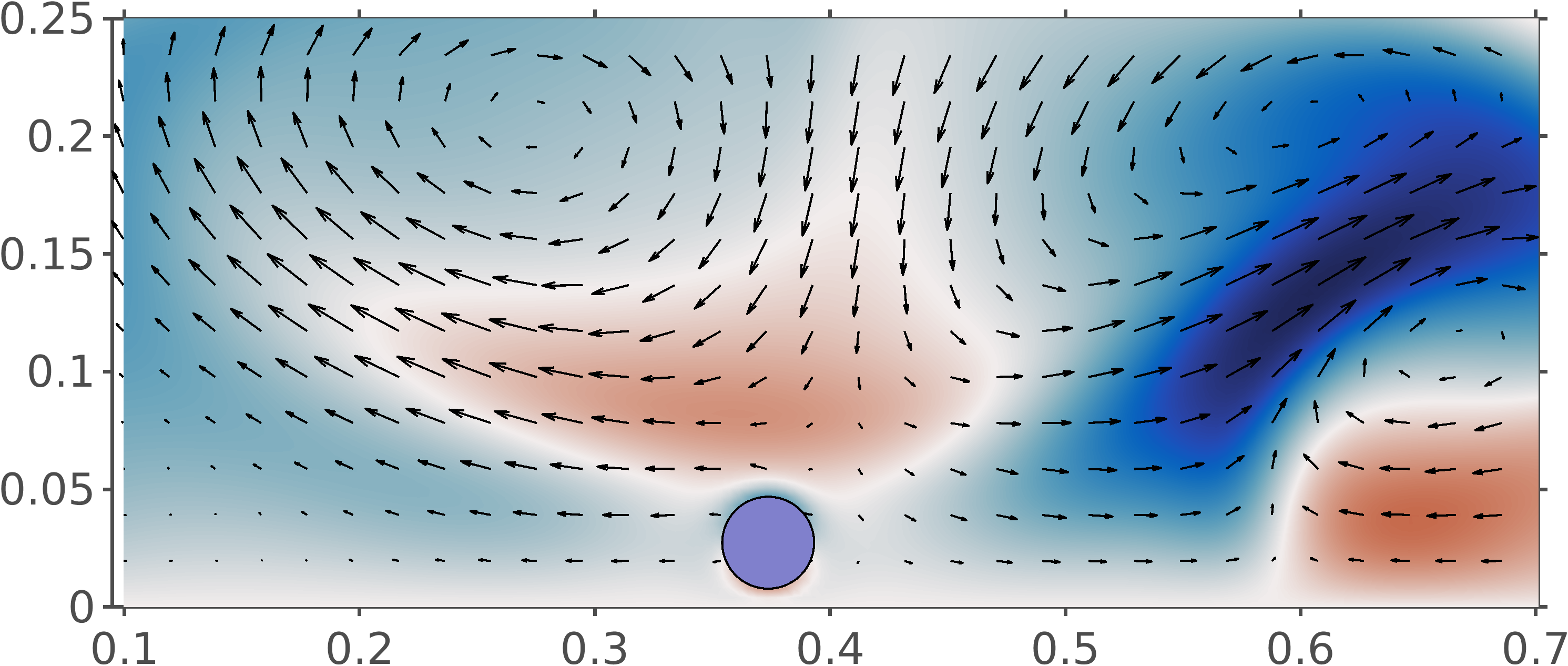}
         \end{minipage}
         \hspace{1ex} 
         \begin{minipage}{2ex}
           \rotatebox{90}
           {\small \hspace{5ex} $y/h$}
         \end{minipage}
         \begin{minipage}{.46\linewidth} 
		       \raggedright{(\textit{c})}
           \includegraphics[width=\linewidth]
           {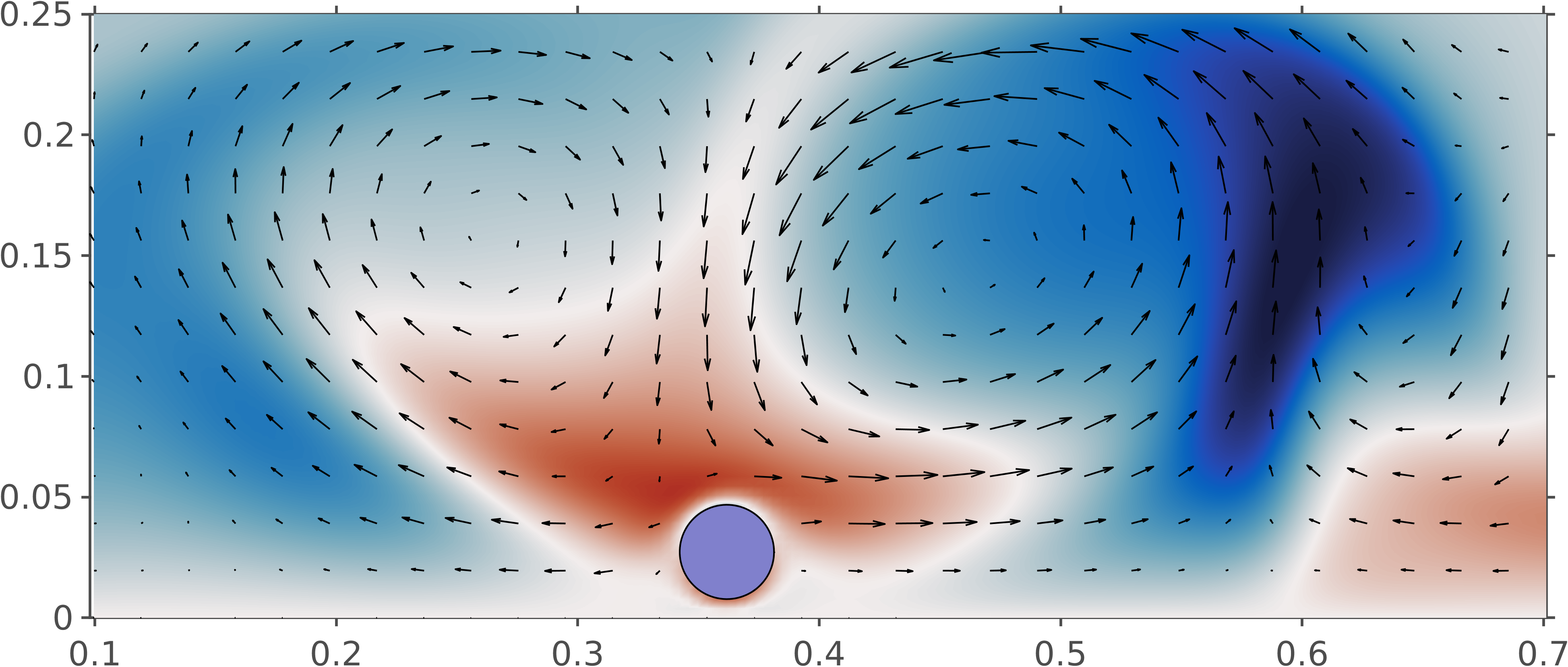}
         \end{minipage}
         \hspace{2ex} 
         \begin{minipage}{2ex}
           \rotatebox{90}
           {\small \hspace{5ex}}
         \end{minipage}
         \begin{minipage}{.45\linewidth} 
           \raggedright{(\textit{d})}
		       \includegraphics[width=\linewidth]
           {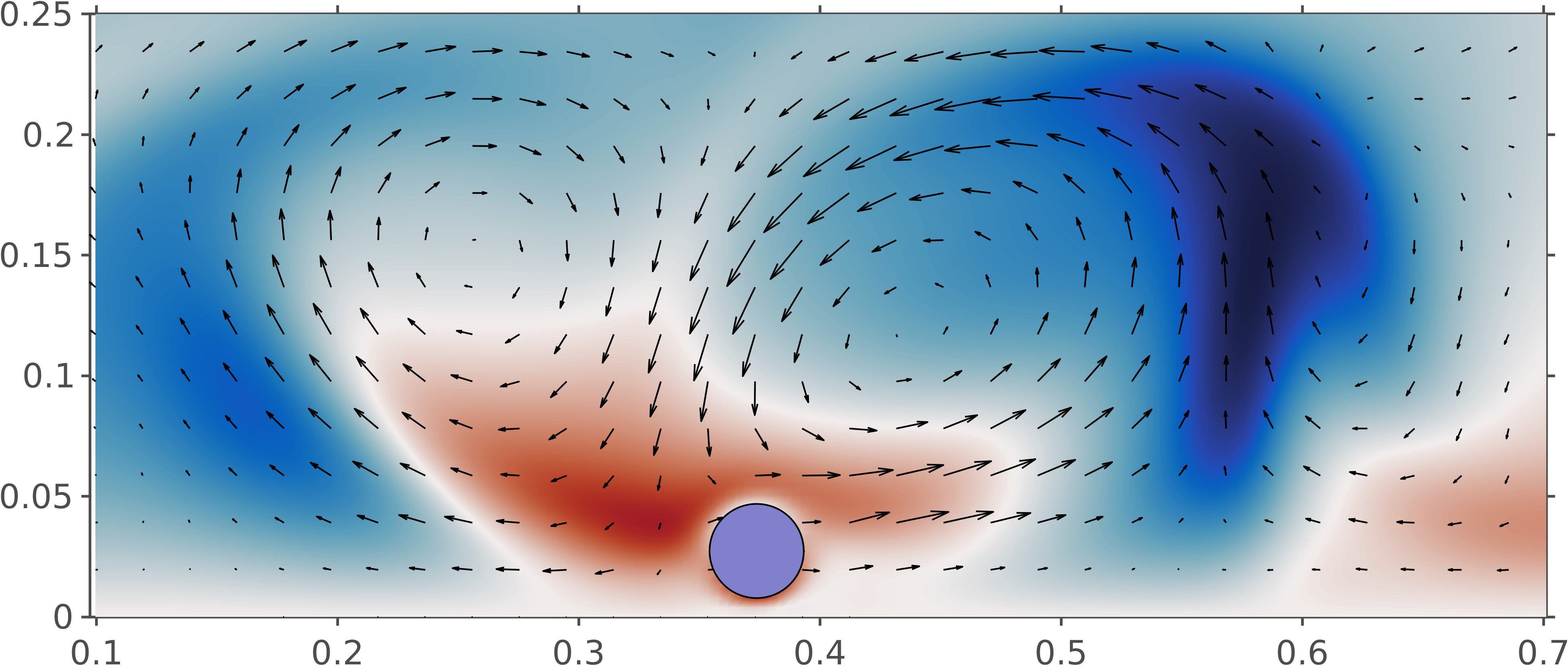}
         \end{minipage}
         \begin{minipage}{2ex}
           \rotatebox{90}
           {\small \hspace{5ex} $y/h$}
         \end{minipage}
         \begin{minipage}{.46\linewidth} 
		       \raggedright{(\textit{e})}
           \includegraphics[width=\linewidth]
           {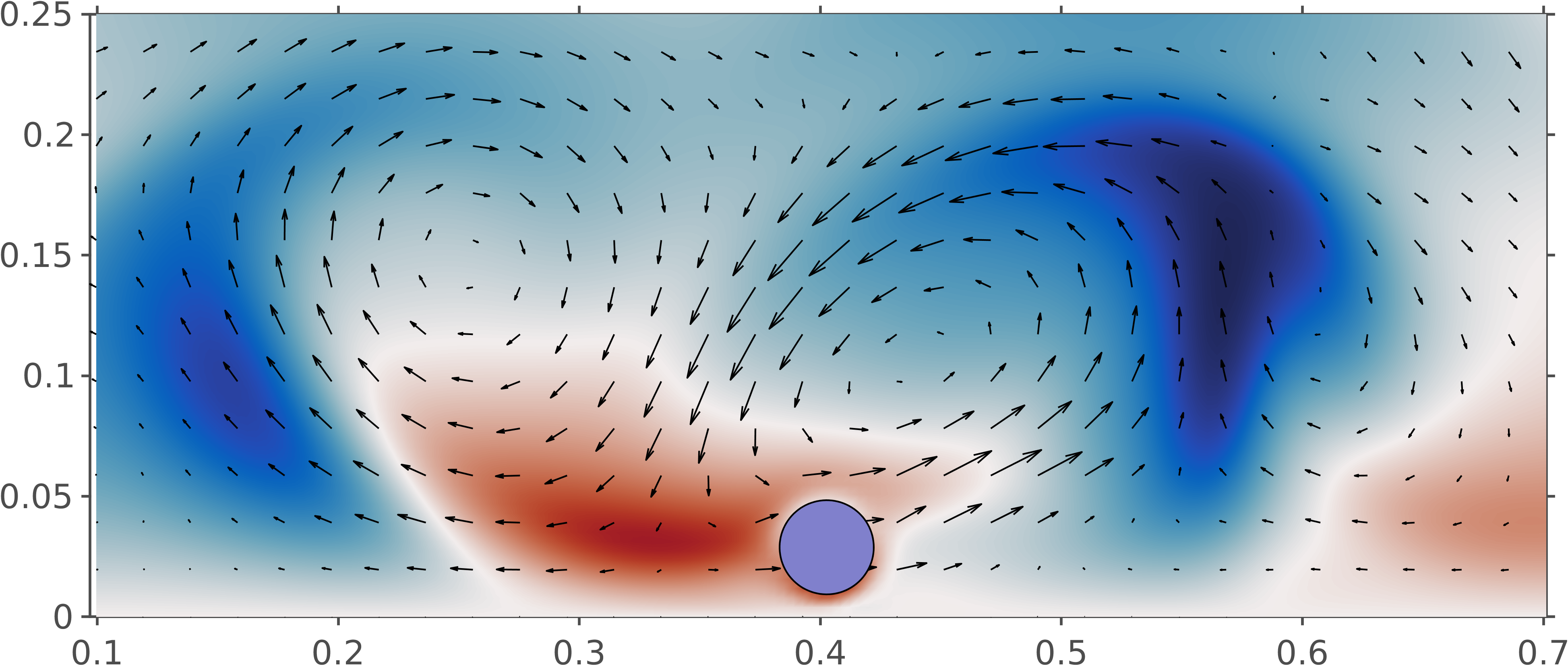}
           \centerline{\small $z/h$}
         \end{minipage}
         \hspace{1ex} 
         \begin{minipage}{2ex}
           \rotatebox{90}
           {\small \hspace{5ex}}
         \end{minipage}
         \begin{minipage}{.46\linewidth} 
           \raggedright{(\textit{f})}
		       \includegraphics[width=\linewidth]
           {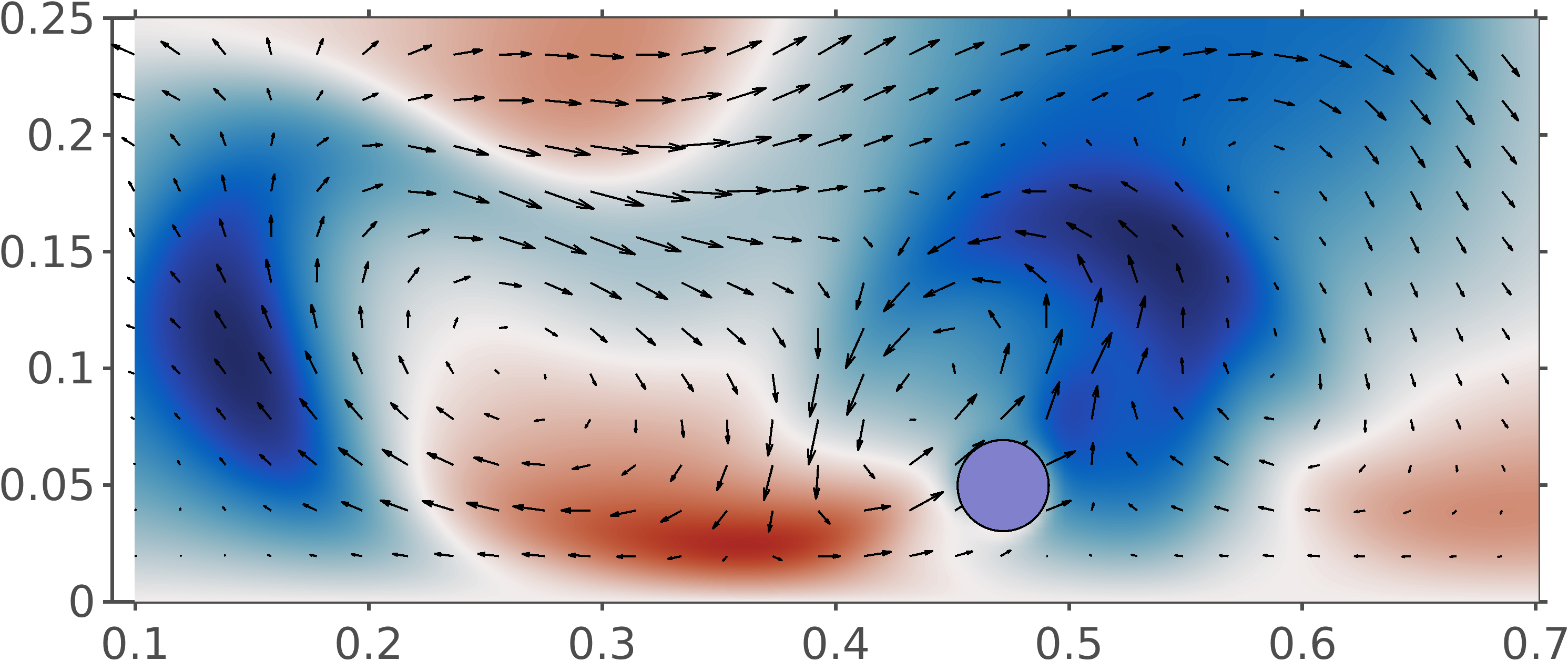}
           \centerline{\small $z/h$}
         \end{minipage}

  	\caption{Time sequence of instantaneous velocity fluctuations of the fluid and the particle in a stream-parallel plane at $x=x_{\mathrm{p}}(t_{\mathrm{r}})/h$. The color contours represent the normalized streamwise velocity fluctuations of fluid, $u_{\mathrm{f}}'/u_\tau$. The arrows denote the velocity fluctuations $(v_{\mathrm{f}}', w_{\mathrm{f}}')/u_\tau$. The snapshots show only a small part of the computational domain. (\textit{a}) $t_{\mathrm{r}}/t_{\mathrm{b}} = -1.0$, (\textit{b}) $t_{\mathrm{r}}/t_{\mathrm{b}} = -0.5$, (\textit{c}) $t_{\mathrm{r}}/t_{\mathrm{b}} = 0.0$, (\textit{d}) $t_{\mathrm{r}}/t_{\mathrm{b}} = 0.25$, (\textit{e}) $t_{\mathrm{r}}/t_{\mathrm{b}} = 0.5$, (\textit{f}) $t_{\mathrm{r}}/t_{\mathrm{b}} = 1.0$.}
\label{fig:snapshots_par20}
\end{figure}

\autoref{fig:snapshots_par20} shows the time evolution of instantaneous velocity fluctuations in a cross-stream slice through a particle from the unconstrained case during an entrainment event. At time $t_\mathrm{r}/t_\mathrm{b} = -1.0$ (\autoref{fig:snapshots_par20}\textit{a}), it can be seen that the particle is located inside a relative low-speed region. Subsequently, a strong sweep event, marked by high streamwise velocity moving down towards the wall, is observed to approach the particle. Evidence of a pair of counter-rotating streamwise vortices approaching the particle can be seen in the vector plot in \autoref{fig:snapshots_par20}(\textit{b}, \textit{c}). In the following, the particle is moving along the spanwise direction towards the low-speed region; this spanwise motion is exactly what was observed by \citet{kidanemariam2013direct} as a result of the presence of a quasi-streamwise vortex with its center located just above the particle (i.e.\ at a vertical distance of the order of ten wall units). In the final part of the sequence (\autoref{fig:snapshots_par20}\textit{e},\textit{f}), the particle of interest is lifted up from the solid wall while the nearest streamwise
vortex (in this case, with counter-clockwise rotation) is approaching even further. As observed on the final two graphs, it is tempting to conjecture that the eventual particle lift-off is decisively influenced by the vortex motion in the $(y,z)$-plane. We will get back to this point in the final discussion section. 

The sequence of events outlined above is consistent with the local wall-normal shear evolution shown in \autoref{fig:shear_rate}: when a sweep is approaching, it leads to an increase of the local shear in the vicinity of a particle. This high shear value measured in the particle's vicinity
will prevail as long as the sweep is located somewhat above the particle. On the contrary, when the particle is fully immersed in the high-speed region (such as seen in \autoref{fig:snapshots_par20}\textit{d},\textit{e}), the local wall-normal shear tends to exhibit smaller values, as was seen in \autoref{fig:shear_rate} at corresponding times ($t_\mathrm{r}/t_\mathrm{b}\ge 0.5$). 

The statistical signature of a similar sequence of events can be identified in the ZC case, although a noticeable time delay is evident in \autoref{fig:shear_rate}. In this configuration, the particle spanwise motion is suppressed, and as a result, a nearby quasi-streamwise vortex would be unable to displace the particle along the spanwise direction. Assuming that the approaching sweep remains in the close vicinity of the particle for a longer duration, a delayed decrease in the local shear rate (\autoref{fig:shear_rate}) is observed. An evaluation of $\langle S_{\mathrm{g}} \rangle$ indicates that the higher magnitude of $\langle S_{\mathrm{g}} \rangle$ (refer to \autoref{tab:frequencies_table}) may be induced due to particles not preferentially accumulating in low-speed regions but also sampling other fluid regions, including high-speed streaks. This observation explains the reduced relative magnitude of $S_{\mathrm{l}}^{\mathrm{ZC}}$ seen in \autoref{fig:shear_rate}.

\subsection{Presence and proximity of vortices}
\label{sec:presence_vortex}

\begin{figure}
\raggedleft
    \begin{tikzpicture}
        \node[anchor=south west,inner sep=0] (fig) at (0,0) 
            {\includegraphics[width=0.95\linewidth]{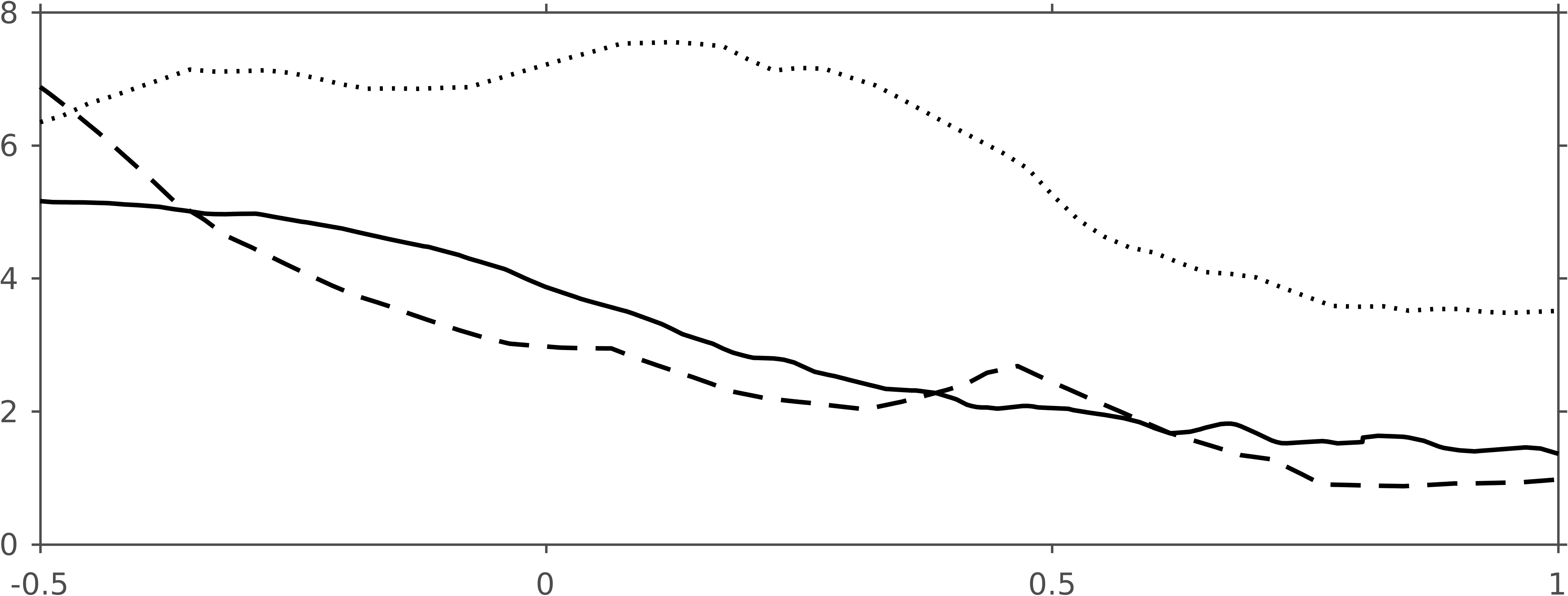}};
        \node at ($(fig.south west)!0.5!(fig.south east)$) [below=3mm, anchor=center, xshift=3.5mm] {$t_{\mathrm{r}}/t_{\mathrm{b}}$};
        \node[rotate=90, anchor=center, yshift=3.5mm, xshift=2.5mm] at ($(fig.south west)!0.5!(fig.north west)$) {$\langle d_{\boldsymbol{\omega}_{\mathrm{f}}} \rangle / D_{\mathrm{p}}$};
    \end{tikzpicture}
    \caption{Time evolution of the distance between the entrained particle center and the closest vortex averaged over all corresponding entrainment events for UN (solid line), RC (dashed line), and ZC (dotted line) cases.}
    \label{fig:vortex_dist}
\end{figure}

The presence of a pair of counter-rotating streamwise vortices in the vicinity of a particle during its entrainment has been observed in \autoref{fig:snapshots_par20}. In order to further investigate the action and relevance of vortices to particle entrainment, the analysis described in \S~\ref{sec:action_of_cortices} was carried out. \autoref{fig:vortex_dist} shows the time evolution of the distance from the center of the particle to the nearest vortex $d_{\boldsymbol{\omega_\mathrm{f}}}$, averaged over the corresponding entrainment events and normalized by the diameter of the particle $D_\mathrm{p}$. The global distance values $\langle d_{\boldsymbol{\omega}_{\mathrm{f}}\mathrm{g}} \rangle/D_\mathrm{p}$ computed over the particles that did not lift off are given in \autoref{tab:frequencies_table}. The data shown in \autoref{fig:vortex_dist} demonstrate that intense vortical structures closely approach the particles before the relevant time interval. After $t_\mathrm{r}=0.25t_\mathrm{b}$, the particles are no longer in contact with the wall, $\langle d_{\boldsymbol{\omega_\mathrm{f}}} \rangle$ continues to decrease, indicating that an intense vortical structure remains in the close proximity of the particles throughout the entrainment process. Considering the geometry of typical buffer-layer structures, this is consistent with a quasi-streamwise vortex of extent of ${\cal O}(100)$ wall units passing by the comparably slow particle.
This result is consistent with the interpretation presented in \S~\ref{sec:shear_rate}, which suggests that particle lift-off is sustained by vortex motion. 

We have seen above that the entrainment process of spanwise constrained particles has a larger duration (cf. the values of $\langle t_\mathrm{rel} \rangle$ given in \autoref{tab:frequencies_table}). Here, we also observe that the approach of the nearest vortex is delayed in case ZC as compared to case UN. Furthermore, the minimum distance is on average larger in the former case.

\section{Summary and discussion}
\label{sec:conclusion}

In the present work, we have numerically studied the lift-off of finite-size heavy particles in a turbulent horizontal open channel flow over a smooth surface under gravitational acceleration. Particles with diameter $D^+_\mathrm{p} = 7.21$ are released in the flow near the wall. Particles were observed to lift off from the wall only rarely, suggesting that such events are due to the action of intense coherent structures. To study the mechanism of lift-off, we first define a threshold based on the wall-normal coordinate to identify entrained particles and select the corresponding time interval that captures the relevant flow dynamics. 

We have analyzed the contributions of hydrodynamic forces, decomposed into drag and lift, to the vertical acceleration using a projection method based on relative velocity (which by itself is not uniquely defined for finite-size particles). The analysis shows that the drag force contribution to particle entrainment is relatively small, while the lift force is the dominant contributor according to this decomposition. Further decomposition of the hydrodynamic forces into pressure and viscous stress components reveals that both contribute almost equally to the total wall-normal force acting on the particle. 

To investigate how the wall-normal force is generated, the local shear rate and the presence of vortices around the particles are analyzed. The local shear rate has been observed to increase before the particle's wall-normal motion begins and to reach its peak shortly thereafter. At the same time, conditional averaging shows that a quasi-streamwise vortex is approaching the particle, as indicated by the analysis of the distance to the nearest vortex. 

We have observed that suppressing particle rotation does not significantly affect any of the observables. This is true both for the global statistics, such as the entrainment frequency (cf. \autoref{tab:frequencies_table}) and for time-resolved conditional averages (such as the force decompositions, the local shear rate, and the distance to the nearest vortex shown in 
Figures~\ref{fig:lift_drag_avg}--\ref{fig:vortex_dist}). From this observation, we can conclude that particle rotation is not an important ingredient of the entrainment mechanism, neither qualitatively nor quantitatively.

Secondly, we have found that suppressing the particle motion in the spanwise direction (while still allowing for streamwise, wall-normal, and full rotational motion) does significantly affect a number of quantities: 
it doubles the average entrainment frequency,
it slows down the lift-off process,
it increases the average (unconditioned) wall-normal shear experienced by particles adjacent to the wall-plane, and it leads to the distance to the nearest vortical structure during entrainment being on average larger than in the unconstrained case. From the comparison between the unconstrained data and the spanwise-motion-constrained data, we can directly conclude that the particle location with respect to the flow structures matters quite sensitively. The fact that unconstrained particles do not sample the flow field uniformly \citep{kidanemariam2013direct} has a significant quantitative influence on the entrainment process.

In order to bring the discussion of a typical entrainment event together, we simultaneously consider the temporal evolution of several key quantities for a single (representative) event pertaining to an unconstrained particle in \autoref{fig:5plots}. An alternative animated sequence is provided in the supplementary material available in \citet{uhlmann:26a-data}. It can be seen that the build-up of a positive wall-normal hydrodynamic force acting on the particle is preceded by a high-shear event (with the peak reaching here nearly twice the global average and on average approximately 1.7 times the global average value). This increase in wall-normal shear around the particle is caused by the arrival of fast-moving fluid above the particle of interest (i.e.\ by a sweep event, cf. \autoref{fig:snapshots_par20}). Simultaneously with the increase in shear occurs the gradual approach of an intense vortical structure, during which distance decreases from an average value of roughly 16 particle diameters (equivalent to approximately 100 wall units, cf. \autoref{tab:frequencies_table}) down to a minimum of roughly 10 wall units. We note from the temporal evolution that while the shear-rate already decreases before the particle fully detaches from the wall, the distance to the nearest vortex still keeps decreasing. Note that we have observed multiple non-entrainment events in which only one of the above two circumstances (high shear-rate, proximity of an intense vortex) occur, but not both. This additional observation of non-entrainment, which merits further quantitative and systematic study in the future, indicates that the combination of these two ingredients (shear and nearby vortex) is necessary in order to lead to an actual entrainment of the particle into the bulk flow at the present parameter point. 

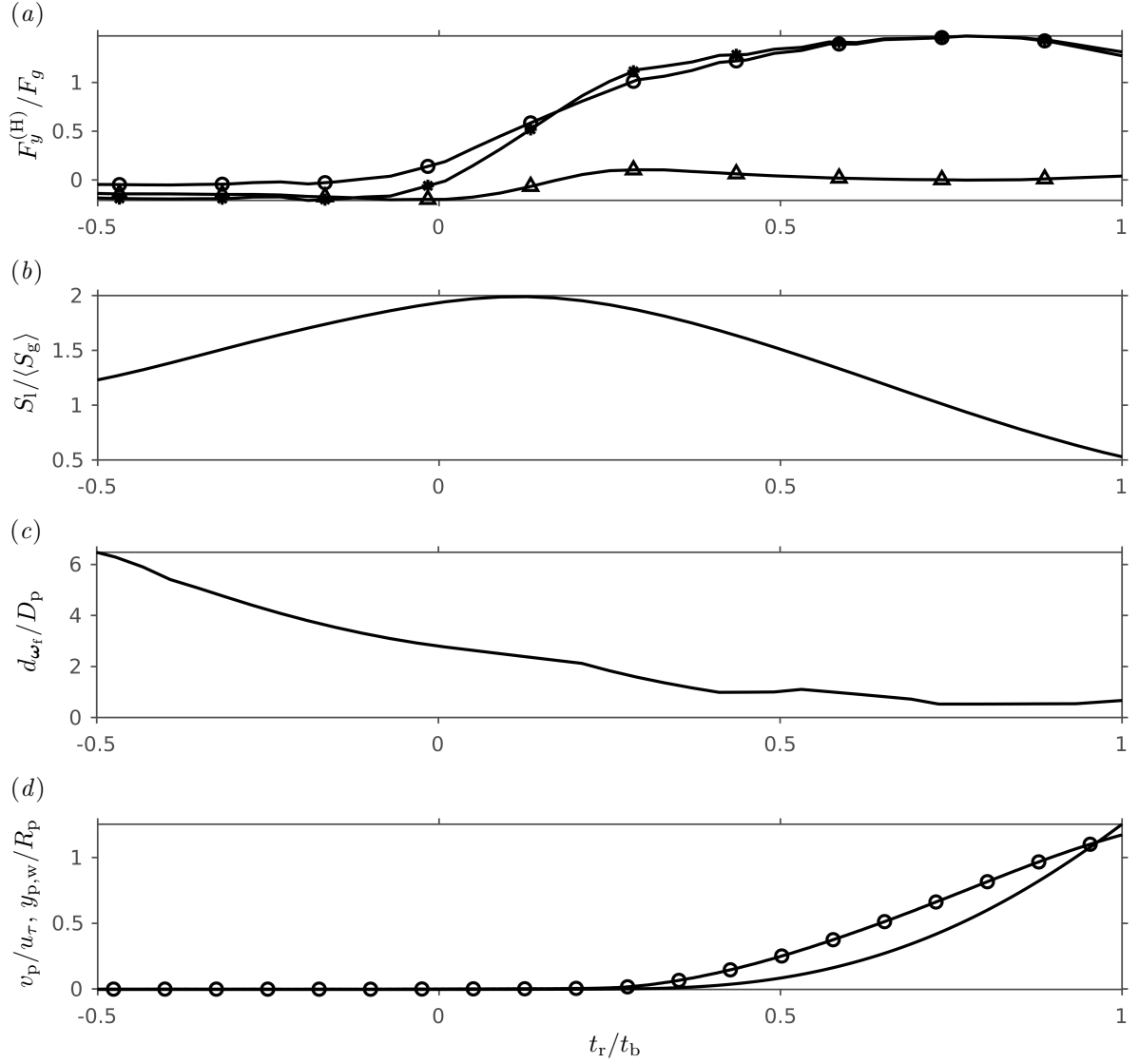
\begin{figure}[!t]
\raggedleft
\vspace{2mm}
    \begin{tikzpicture}
        \node[anchor=south west,inner sep=0] (fig) at (0,0) 
            {\includegraphics[width=0.93\linewidth]{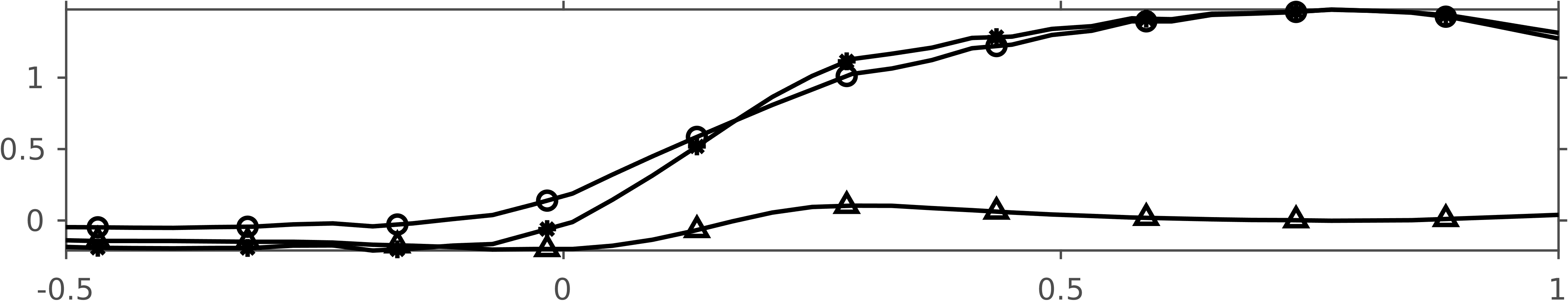}};
        \node[rotate=90, anchor=center, yshift=3.5mm, xshift=2.5mm] at ($(fig.south west)!0.5!(fig.north west)$) {$ F_y^{(\mathrm{H})} /F_g$};
        \node[anchor=north west, xshift=-7mm, yshift=5mm] at (fig.north west) {(\textit{a})};
    \end{tikzpicture}
    
    \vspace{2mm}
    
    \begin{tikzpicture}
        \node[anchor=south west,inner sep=0] (fig) at (0,0) 
            {\includegraphics[width=0.93\linewidth]{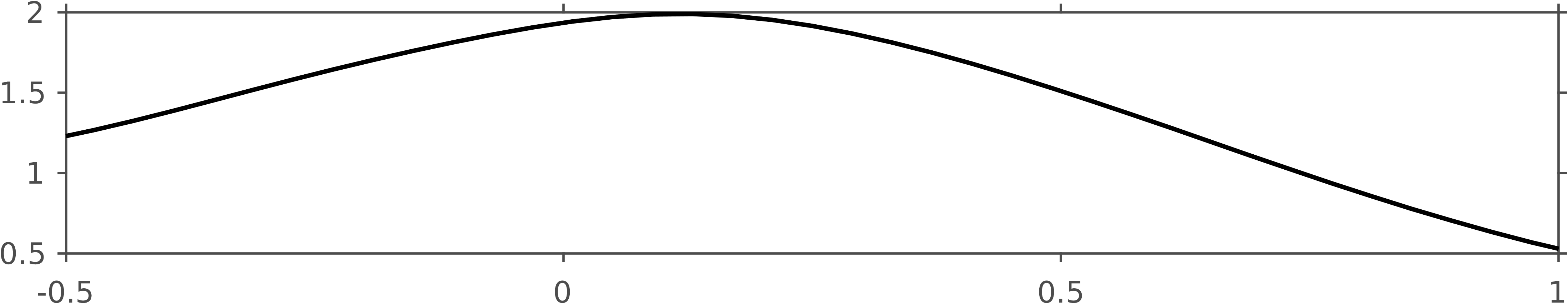}};
        \node[rotate=90, anchor=center, yshift=3.5mm, xshift=2.5mm] at ($(fig.south west)!0.5!(fig.north west)$) {$ S_{\mathrm{l}} / \langle S_{\mathrm{g}} \rangle$};
        \node[anchor=north west, xshift=-7mm, yshift=5mm] at (fig.north west) {(\textit{b})};
    \end{tikzpicture}
    
    \vspace{2mm}
    
    \begin{tikzpicture}
        \node[anchor=south west,inner sep=0] (fig) at (0,0) 
            {\includegraphics[width=0.915\linewidth]{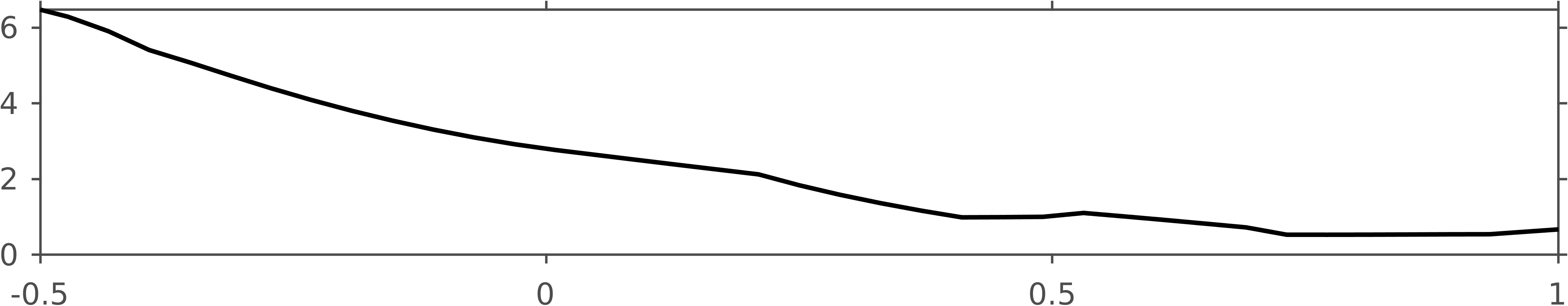}};
        \node[rotate=90, anchor=center, yshift=5.5mm, xshift=2.5mm] at ($(fig.south west)!0.5!(fig.north west)$) {$d_{\boldsymbol{\omega}_{\mathrm{f}}} / D_{\mathrm{p}}$};
        \node[anchor=north west, xshift=-9.5mm, yshift=5mm] at (fig.north west) {(\textit{c})};
    \end{tikzpicture}
    
    \vspace{2mm}
    
    \begin{tikzpicture}
        \node[anchor=south west,inner sep=0] (fig) at (0,0) 
            {\includegraphics[width=0.93\linewidth]{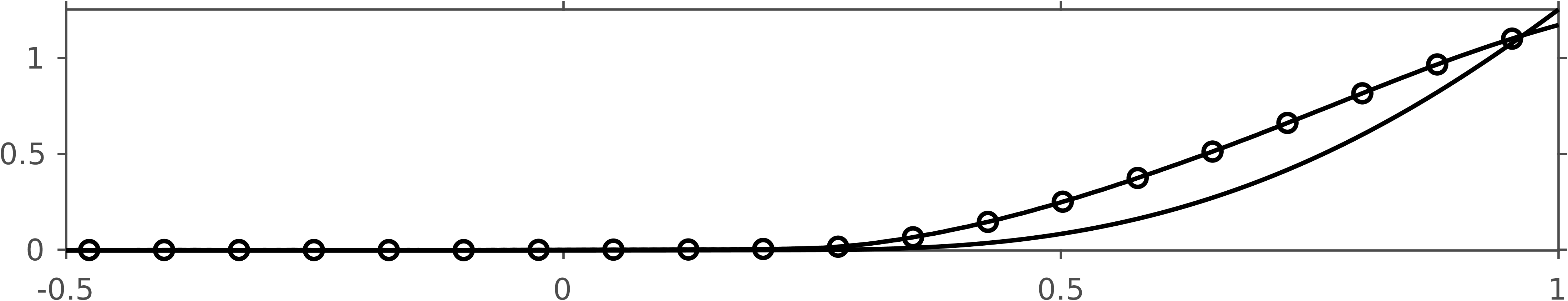}};
        \node[rotate=90, anchor=center, yshift=3.5mm, xshift=2.5mm] at ($(fig.south west)!0.5!(fig.north west)$) {$v_{\mathrm{p}}/u_{\tau}$, $y_{\mathrm{p,w}}/R_{\mathrm{p}}$};
        \node at ($(fig.south west)!0.5!(fig.south east)$) [below=3mm, anchor=center, xshift=3.5mm] {$t_{\mathrm{r}}/t_{\mathrm{b}}$};
        \node[anchor=north west, xshift=-7mm, yshift=7mm] at (fig.north west) {(\textit{d})};
    \end{tikzpicture}
    \caption{Time evolution of different parameters related to one specific (representative) particle during entrainment. (\textit{a}) wall-normal force $F_y^{(\mathrm{H})}$ (\textasteriskcentered) with lift $F_{y,\mathrm{lift}}^{(\mathrm{H})}$ (\(\circ\)) and drag $F_{y,\mathrm{drag}}^{(\mathrm{H})}$ (\(\scalebox{0.7}{$\triangle$}\)) contributions. (\textit{b}) local wall-normal shear around the particle $S_{\mathrm{l}}$. (\textit{c}) distance to the nearest intense vortical structure. (\textit{d}) wall-normal particle velocity $v_{\mathrm{p}}$ (\(\circ\)) and wall-normal distance $y_{\mathrm{p,w}}$ (-).}
    \label{fig:5plots}
\end{figure}

From the above discussion, it appears that the two essential elements for the entrainment of a particle from a smooth wall at marginal Shields number values are: (i) a high wall-normal shear rate, and (ii) the proximity of an intense quasi-streamwise vortex. Yet the question remains as to how the wall-normal force that lifts the particle off the wall is being generated in detail. According to our discussion, we might attempt to simplify the situation to a particle translating along a wall in a simple shear flow, featuring a single quasi-streamwise vortex (or a pair of counter-rotating vortices). Unfortunately, this highly idealized configuration is already too complicated for it to be amenable to an analytic treatment. Even empirical force models, such as those which are employed in the context of a point-particle ansatz (cf.\ discussion in \S~\ref{sec:introduction}), are typically derived from data in even simpler configurations. More specifically, parametric studies of this type feature either a particle translating in a wall-bounded linear shear flow \citep[e.g.][]{shi2021drag}, or a particle translating in an unbounded vortical flow \citep{bagchi2002shearvortex}. Therefore, it is not feasible at the current time to estimate the respective relative contributions of shear-induced and vortex-induced lift to the entrainment process from theoretical or semi-empirical considerations.

In order to close the above-mentioned knowledge gap, additional work is needed to establish causality. In particular, further studies employing simplified configurations, targeted interventions, or conditional analyses could be used to systematically probe the mechanisms underlying particle entrainment. Such studies would help to verify or falsify the above outlined scenario and to disentangle the respective roles of local shear and vortical structures.

As a further perspective on the present study, let us point out that the amount of samples analyzed so far is rather marginal. This point should be addressed in the future. Based on the currently developed methodology and workflow, carrying out this task will be straightforward.\

Finally, it should be mentioned that it will be of interest to vary the physical parameters of the problem in order to establish the scaling of the investigated mechanism. Apart from the various particle-related parameters ($D^+$, $\rho_p/\rho_f$, $\Phi_\mathrm{s}$), it will be of particular interest to investigate the effect of large-scale and very large-scale flow structures upon the particle dynamics. In open channel flow, these large flow structures have been shown to arise at lower Reynolds numbers than in the corresponding closed channel flow, while also being more strongly correlated in space \citep{bauer2025far}.

\section*{Acknowledgments}

The current work was supported by the German Research Foundation (DFG) through grant UH242/15-1, and in part by the European Research Council under the Caust grant ERC-AdG-101018287. The simulations were performed on the supercomputer bwUniCluster funded by the Ministry of Science, Research and the Arts Baden-Württemberg and the Universities of the State of Baden-Württemberg.

\appendix
\section{Effect of the choice of the shell radius in particle-conditioned averaging upon force components}
\label{sec:appendix_rs}

\begin{figure}[t]
\raggedleft
    \begin{tikzpicture}
        \node[anchor=south west,inner sep=0] (fig) at (0,0) 
            {\includegraphics[width=0.95\linewidth]{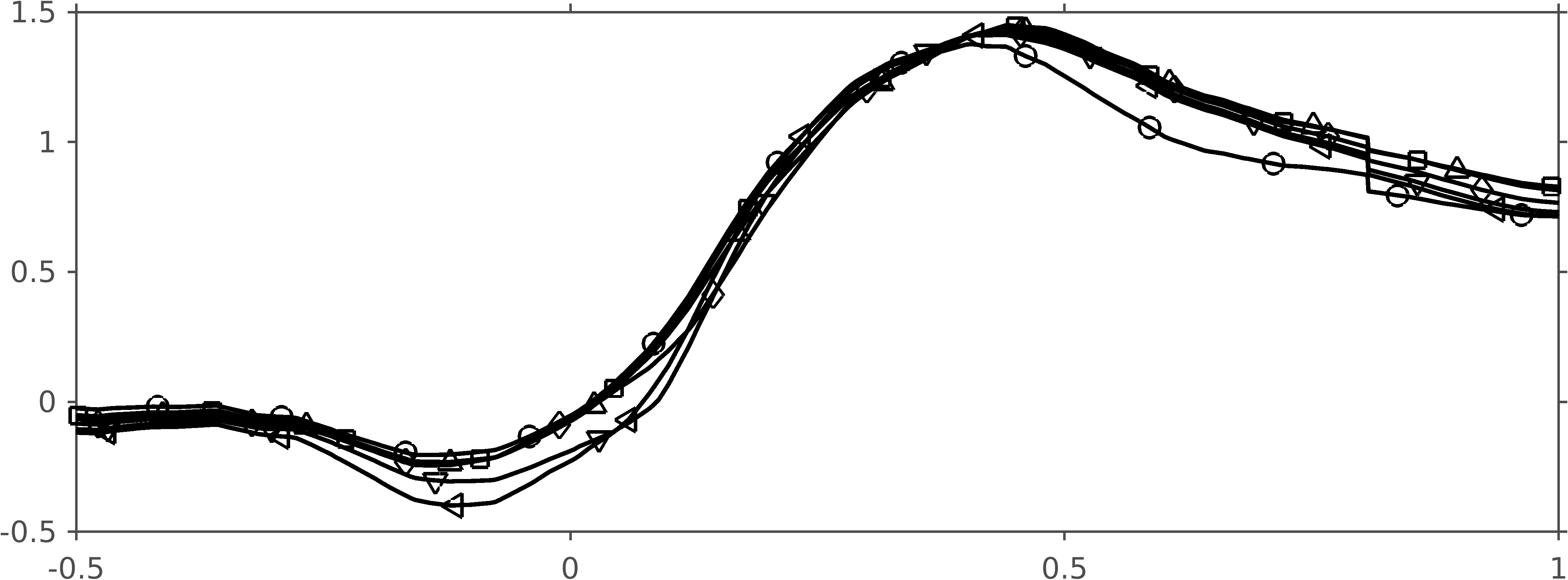}};
        \node at ($(fig.south west)!0.5!(fig.south east)$) [below=3mm, anchor=center, xshift=3.5mm] {$t_{\mathrm{r}}/t_{\mathrm{b}}$};
        \node[rotate=90, anchor=center, yshift=3.5mm, xshift=2.5mm] at ($(fig.south west)!0.5!(fig.north west)$) {$\langle F_{y,\mathrm{lift}}^{(\mathrm{H})} \rangle /F_g$};
    \end{tikzpicture}
    \caption{Time evolution of lift contribution of hydrodynamic force in wall-normal direction $F_{y,\mathrm{lift}}^{(\mathrm{H})}$ evaluated with different values of the averaging shell radius $R_s$. The data is averaged over all corresponding entrainment events for the UN case. $R_s = 2R_\mathrm{p}$ ($\circ$), $R_s = 2.5R_\mathrm{p}$ ($\square$), $R_s = 3R_\mathrm{p}$ ($\triangle$), $R_s = 3.5R_\mathrm{p}$ ($\diamond$), $R_s = 4R_\mathrm{p}$ ($\triangledown$), $R_s = 4.5R_\mathrm{p}$ ($\triangleleft$).}
    \label{fig:lift_R}
\end{figure}

A sensitivity analysis was performed to evaluate the effect of the radius $R_s$ of the spherical averaging surface (cf. \autoref{fig:shell_and_forces}\textit{a}) on the force decomposition. Here, only the lift contribution for the UN case is shown for clarity. \autoref{fig:lift_R} represents the sensitivity analysis of the contribution of the lift force to the total hydrodynamic force in the wall-normal direction with respect to the $R_s$ value. As we observe from \autoref{fig:lift_R}, varying $R_s$ does not lead to significant changes in the computed force contribution. Despite small differences, all values of $R_s$ in the range $2R_\mathrm{p}$ to $4.5R_\mathrm{p}$ lead to a lift force being the dominant force in the entrainment of the particle. The results confirm that the chosen value of $R_s = 3R_\mathrm{p}$ lies in a range where the force contribution is only little sensitive to $R_S$.

\bibliographystyle{bst/iop_confStyleMU}


\bibliography{
  entrainment_bibliography.bib
}

@article{dwivedi2011flow,
  title={Flow structures and hydrodynamic force during sediment entrainment},
  author={Dwivedi, A and Melville, B W and Shamseldin, A Y and Guha, T K},
  journal={Water Resour. Res.},
  volume={47},
  number={1},
  doi = {https://doi.org/10.1029/2010WR009089},
  year={2011},
  publisher={Wiley Online Library}
}

@article{dwivedipressure,
author = {Dwivedi, A and Melville, B and Shamseldin, A Y},
title = {Hydrodynamic Forces Generated on a Spherical Sediment Particle during Entrainment},
journal = {J. Hydraul. Eng.},
volume = {136},
number = {10},
pages = {756-769},
year = {2010},
doi = {10.1061/(ASCE)HY.1943-7900.0000247},
}

@article{bagchi2002shearvortex,
  title={Shear versus vortex-induced lift force on a rigid sphere at moderate Re},
  author={Bagchi, P and Balachandar, S},
  journal={J. Fluid Mech.},
  volume={473},
  pages={379-388},
  year={2002},
  doi={10.1017/S0022112002002628},
  publisher={Cambridge University Press}
}

@article{kidanemariam2013direct,
  title={Direct numerical simulation of horizontal open channel flow with finite-size, heavy particles at low solid volume fraction},
  author={Kidanemariam, A G and Chan-Braun, C and Doychev, T and Uhlmann, M},
  journal={New J. Phys.},
  volume={15},
  number={2},
  pages={025031},
  year={2013},
  doi = {10.1088/1367-2630/15/2/025031},
  publisher={IOP Publishing}
}

@article{ninto1996experiments,
  title={Experiments on particle—turbulence interactions in the near--wall region of an open channel flow: implications for sediment transport},
  author={Niño, Y and Garcia, M H},
  journal={J. Fluid Mech.},
  volume={326},
  pages={285-319},
  year={1996},
  doi={10.1017/S0022112096008324},
  publisher={Cambridge University Press}
}

@article{suzuki2000simultaneous,
  title={Simultaneous measurement of fluid and dispersed phases in a particle-laden turbulent channel flow with the aid of 3-D PTV},
  author={Suzuki, Y and Ikenoya, M and Kasagi, N},
  journal={Exp. Fluids.},
  volume={29},
  pages={S185-S193},
  year={2000},
  doi={10.1007/s003480070020},
  publisher={Springer}
}

@article{pestana2020can,
  title={Can preferential concentration of finite-size particles in plane Couette turbulence be reproduced with the aid of equilibrium solutions?},
  author={Pestana, T and Uhlmann, M and Kawahara, G},
  journal={Phys. Rev. Fluids},
  volume={5},
  number={3},
  pages={034305},
  year={2020},
  doi = {10.1103/PhysRevFluids.5.034305},
  publisher={APS}
}

@article{peng2024preferential,
  title={Preferential accumulation of finite-size particles in near-wall streaks},
  author={Peng, C and Wang, L P and Chen, S},
  journal={J. Fluid Mech.},
  volume={980},
  pages={A38},
  year={2024},
  doi={10.1017/jfm.2024.41},
  publisher={Cambridge University Press}
}

@TechReport{shields:36,
  author =        {Shields, A},
  title =         {Anwendungen der {{\"{A}}}hnlichkeitsmechanik und der
                   {T}urbulenzforschung auf die {G}eschiebebewegung},
  institution =  {Mitteilungen der Preu{\ss}ischen Versuchsanstalt
                   f\"ur Wasser- Erd- und Schiffbau},
  year =          {1936},
  number =        {26},
  address =       {Berlin, Germany}
}

@article{HENRY20231,
title = {Particle resuspension: Challenges and perspectives for future models},
journal = {Physics Reports},
volume = {1007},
pages = {1-98},
year = {2023},
doi = {https://doi.org/10.1016/j.physrep.2022.12.005},
author = {Henry, C and Minier, J P and Brambilla, S},
}

@article{loth2008lift,
  title={Lift of a spherical particle subject to vorticity and/or spin},
  author={Loth, E},
  journal={AIAA J.},
  volume={46},
  number={4},
  pages={801-809},
  year={2008},
  doi = {10.2514/1.29159},
}

@article{saffman1965lift,
  title={The lift on a small sphere in a slow shear flow},
  author={Saffman, P G},
  journal={J. Fluid Mech.},
  volume={22},
  number={2},
  pages={385-400},
  year={1965},
  doi={10.1017/s0022112065000824},
  publisher={Cambridge University Press}
}

@article{patankar2001lift,
  title={Lift-off of a single particle in Newtonian and viscoelastic fluids by direct numerical simulation},
  author={Patankar, N A and Huang, P Y and Ko, T and Joseph, D D},
  journal={J. Fluid Mech.},
  volume={438},
  pages={67-100},
  year={2001},
  doi={10.1017/S0022112001004104},
  publisher={Cambridge University Press}
}

@article{joseph2002slip,
  title={Slip velocity and lift},
  author={Joseph, D D and Ocando, D},
  journal={J. Fluid Mech.},
  volume={454},
  pages={263-286},
  year={2002},
  doi={10.1017/S0022112001007145},
  publisher={Cambridge University Press}
}

@article{uhlmann2005immersed,
  title={An immersed boundary method with direct forcing for the simulation of particulate flows},
  author={Uhlmann, M},
  journal={J. Comput. Phys.},
  volume={209},
  number={2},
  pages={448-476},
  year={2005},
  doi = {https://doi.org/10.1016/j.jcp.2005.03.017},
  publisher={Elsevier}
}

@article{li2019fully,
  title={Fully resolved simulations of a stationary finite-sized particle in wall turbulence over a rough bed},
  author={Li, X and Balachandar, S and Lee, H and Bai, B},
  journal={Phys. Rev. Fluids},
  volume={4},
  number={9},
  pages={094302},
  year={2019},
  doi = {10.1103/PhysRevFluids.4.094302},
  publisher={APS}
}

@article{cisse2013slipping,
  title={Slipping motion of large neutrally buoyant particles in turbulence},
  author={Cisse, M and Homann, H and Bec, J},
  journal={J. Fluid Mech.},
  volume={735},
  pages={R1},
  year={2013},
  doi={10.1017/jfm.2013.490},
  publisher={Cambridge University Press}
}

@article{zeng2005wall,
  title={Wall-induced forces on a rigid sphere at finite Reynolds number},
  author={Zeng, L and Balachandar, S and Fischer, P},
  journal={J. Fluid Mech.},
  volume={536},
  pages={1-25},
  year={2005},
  doi={10.1017/S0022112005004738},
  publisher={Cambridge University Press}
}

@article{zeng2009forces,
  title={Forces on a finite-sized particle located close to a wall in a linear shear flow},
  author={Zeng, L and Najjar, F and Balachandar, S and Fischer, P},
  journal={Physics of fluids},
  volume={21},
  number={3},
  pages = {033302},
  year={2009},
  doi = {10.1063/1.3082232},
  publisher={AIP Publishing}
}

@article{lee2010drag,
  title={Drag and lift forces on a spherical particle moving on a wall in a shear flow at finite Re},
  author={Lee, H and Balachandar, S},
  journal={J. Fluid Mech.},
  volume={657},
  pages={89-125},
  year={2010},
  doi={10.1017/S0022112010001382},
  publisher={Cambridge University Press}
}

@article{shi2021drag,
  title={Drag and lift forces on a rigid sphere immersed in a wall-bounded linear shear flow},
  author={Shi, P and Rzehak, R and Lucas, D and Magnaudet, J},
  journal={Phys. Rev. Fluids},
  volume={6},
  number={10},
  pages={104309},
  year={2021},
  doi = {10.1103/PhysRevFluids.6.104309},
  publisher={APS}
}

@article{maxey1983equation,
  title={Equation of motion for a small rigid sphere in a nonuniform flow},
  author={Maxey, M R and Riley, J J},
  journal={The Physics of Fluids},
  volume={26},
  number={4},
  pages={883-889},
  year={1983},
  doi = {10.1063/1.864230},
  publisher={AIP Publishing}
}

@article{gatignol1983faxen,
  title={The Fax{\'e}n formulae for a rigid particle in an unsteady non-uniform Stokes flow},
  author={Gatignol, R},
  year={1983},
  journal={J. Méc. Théor. Appl.},
  pages={143--160},
  volume={2},
  number={2}
}

@article{m2016point,
  title={Point-particle DNS and LES of particle-laden turbulent flow-a state-of-the-art review},
  author={Kuerten, J G},
  journal={Flow Turbul. Combust.},
  volume={97},
  number={3},
  pages={689-713},
  year={2016},
  doi = {10.1007/s10494-016-9765-y},
  publisher={Springer}
}

@article{mclaughlin1991inertial,
  title={Inertial migration of a small sphere in linear shear flows},
  author={McLaughlin, J B},
  journal={J. Fluid Mech.},
  volume={224},
  pages={261-274},
  year={1991},
  doi={10.1017/S0022112091001751},
  publisher={Cambridge University Press}
}

@article{mei1994note,
  title={A note on the history force on a spherical bubble at finite Reynolds number},
  author={Mei, R and Klausner, J F and Lawrence, C J},
  journal={Phys. Fluids},
  volume={6},
  number={1},
  pages={418-420},
  year={1994},
  doi = {10.1063/1.868039},
  publisher={American Institute of Physics}
}

@article{rubinow1961transverse,
  title={The transverse force on a spinning sphere moving in a viscous fluid},
  author={Rubinow, S I and Keller, J B},
  journal={J. Fluid Mech.},
  volume={11},
  number={3},
  pages={447-459},
  year={1961},
  doi={10.1017/S0022112061000640},
  publisher={Cambridge University Press}
}

@article{bagchi2002effect,
  title={Effect of free rotation on the motion of a solid sphere in linear shear flow at moderate Re},
  author={Bagchi, P and Balachandar, S},
  journal={Phys. Fluids},
  volume={14},
  number={8},
  pages={2719-2737},
  year={2002},
  doi = {10.1063/1.1487378},
  publisher={American Institute of Physics}
}

@article{shi2019lift,
  title={Lift forces on solid spherical particles in unbounded flows},
  author={Shi, P and Rzehak, R},
  journal={Chem. Eng. Sci.},
  volume={208},
  pages={115145},
  year={2019},
  doi = {https://doi.org/10.1016/j.ces.2019.08.003},
  publisher={Elsevier}
}

@article{soldati2009physics,
  title={Physics and modelling of turbulent particle deposition and entrainment: Review of a systematic study},
  author={Soldati, A and Marchioli, C},
  journal={Int. J. Multiph. Flow},
  volume={35},
  number={9},
  pages={827-839},
  year={2009},
  doi = {https://doi.org/10.1016/j.ijmultiphaseflow.2009.02.016},
  publisher={Elsevier}
}

@article{vowinckel2016entrainment,
  title={Entrainment of single particles in a turbulent open-channel flow: A numerical study},
  author={Vowinckel, B and Jain, R and Kempe, T and Fr{\"o}hlich, J},
  journal={J. Hydraul. Res.},
  volume={54},
  number={2},
  pages={158-171},
  year={2016},
  doi = {10.1080/00221686.2016.1140683},
  publisher={Taylor \& Francis}
}

@article{cameron2020entrainment,
  title={Entrainment of sediment particles by very large-scale motions},
  author={Cameron, S M and Nikora, V I and Witz, M J},
  journal={J. Fluid Mech.},
  volume={888},
  pages={A7},
  year={2020},
  doi={10.1017/jfm.2020.24},
  publisher={Cambridge University Press}
}

@article{yousefi2020single,
  title={Single sediment dynamics in turbulent flow over a porous bed--insights from interface-resolved simulations},
  author={Yousefi, A and Costa, P and Brandt, L},
  journal={J. Fluid Mech.},
  volume={893},
  pages={A24},
  year={2020},
  doi={10.1017/jfm.2020.242},
  publisher={Cambridge University Press}
}

@article{annurev:/content/journals/10.1146/annurev.fl.23.010191.003125,
   author = "Robinson, S K",
   title = "Coherent Motions in the Turbulent Boundary Layer", 
   journal= "Annu. Rev. Fluid Mech.",
   year = "1991",
   volume = "23",
   pages = "601-639",
   publisher = "Annual Reviews",
   doi = "https://doi.org/10.1146/annurev.fl.23.010191.003125",
   type = "Journal Article",
  }

@article{jimenez2018coherent,
  title={Coherent structures in wall-bounded turbulence},
  author={Jim{\'e}nez, J},
  journal={J. Fluid Mech.},
  volume={842},
  pages={P1},
  year={2018},
  doi={10.1017/jfm.2018.144},
  publisher={Cambridge University Press}
}

@InProceedings{hunt1988eddies,
  author =        {Hunt, J C R and Wray, A A and Moin, P},
  title =         {Eddies, streams, and convergence zones in turbulent flows},
  booktitle =     {Proceedings of the Summer Programm},
  pages =         {193-208},
  year =          1988,
  publisher = {(Center for Turbulence Research, Stanford)}}

@article{uhlmann2017clustering,
  title={Clustering and preferential concentration of finite-size particles in forced homogeneous-isotropic turbulence},
  author={Uhlmann, M and Chouippe, A},
  journal={J. Fluid Mech.},
  volume={812},
  pages={991-1023},
  year={2017},
  doi={10.1017/jfm.2016.826},
  publisher={Cambridge University Press}
}

@article{peskin2002immersed,
  title={The immersed boundary method},
  author={Peskin, C S},
  journal={Acta Numer.},
  volume={11},
  pages={479-517},
  year={2002},
  doi={10.1017/S0962492902000077},
  publisher={Cambridge University Press}
}

@article{glowinski1999distributed,
  title={A distributed Lagrange multiplier/fictitious domain method for particulate flows},
  author={Glowinski, R and Pan, T W and Hesla, T I and Joseph, D D},
  journal={ Int. J. Multiph. Flow},
  volume={25},
  number={5},
  pages={755-794},
  year={1999},
  doi = {https://doi.org/10.1016/S0301-9322(98)00048-2},
  publisher={Elsevier}
}

@article{uhlmann2014sedimentation,
  title={Sedimentation of a dilute suspension of rigid spheres at intermediate Galileo numbers: the effect of clustering upon the particle motion},
  author={Uhlmann, M and Doychev, T},
  journal={J. Fluid Mech.},
  volume={752},
  pages={310-348},
  year={2014},
  doi={10.1017/jfm.2014.330},
  publisher={Cambridge University Press}
}

@article{bauer2025far,
  title={How far does the influence of the free surface extend in turbulent open channel flow?},
  author={Bauer, C and Sakai, Y and Uhlmann, M},
  journal={Flow Turbul. Combust.},
  volume={115},
  pages={447–468},
  year={2025},
  doi = {10.1007/s10494-025-00665-4},
  publisher={Springer}
}

@Misc{uhlmann:26a-data,
  author =        {Uhlmann, M and Bzikadze, T},
  title =         {Animation of particle lift-off in turbulent open channel flow based on data from {PR-DNS}},
  howpublished = {KITopen},
  note =          {(Animation)},
  year =          2026,
  doi =          {10.35097/aw3dzkum9ud9r6yf}
}

@article{BILOTTA20082849,
title = {Understanding the influence of suspended solids on water quality and aquatic biota},
journal = {Water Res.},
volume = {42},
number = {12},
pages = {2849-2861},
year = {2008},
doi = {https://doi.org/10.1016/j.watres.2008.03.018},
author = {Bilotta, G S and Brazier, R E},
}

@article{rochmanpollution,
author = {Rochman, C M},
title = {Microplastics research—from sink to source},
journal = {Science},
volume = {360},
number = {6384},
pages = {28-29},
year = {2018},
doi = {10.1126/science.aar7734},
}

@article{https://doi.org/10.1046/j.1365-3091.2003.00551.x,
author = {Niño, Y and Lopez, F and Garcia, M},
title = {Threshold for particle entrainment into suspension},
journal = {Sedimentology},
volume = {50},
number = {2},
pages = {247-263},
doi = {https://doi.org/10.1046/j.1365-3091.2003.00551.x},
year = {2003}
}

\end{document}